\documentclass[manuscript,screen,authorversion,nonacm]{acmart}

\usepackage{multirow}

\AtBeginDocument{%
    \addtolength{\footskip}{2.0\baselineskip}%
    \fancyfoot[L]{\textit{\textbf{Preprint --- do not distribute.}}}%
}

\AtBeginDocument{%
  \providecommand\BibTeX{{%
    \normalfont B\kern-0.5em{\scshape i\kern-0.25em b}\kern-0.8em\TeX}}}

\setcopyright{acmcopyright}
\copyrightyear{2021}
\acmYear{2021}
\acmDOI{10.1145/1122445.1122456}

\begin{document}

\title{Towards a Theory on Architecting for Continuous Deployment}


\author{Breno Bernard Nicolau de França}
\email{breno@ic.unicamp.br}
\orcid{1234-5678-9012}
\affiliation{%
  \institution{Universidade Estadual de Campinas}
  \streetaddress{P.O. Box 6176}
  \city{Campinas}
  \state{São Paulo}
  \country{Brazil}
  \postcode{13083-970}
}

\author{Paulo Sérgio Medeiros dos Santos}
\affiliation{%
  \institution{Universidade Federal do Estado do Rio de Janeiro}
  \city{Rio de Janeiro}
  \state{Rio de Janeiro}
  \country{Brazil}}
\email{pasemes@uniriotec.br}

\author{Santiago Matalonga}
\affiliation{%
  \institution{The University of the West of Scotland}
  \country{Scotland}
}
\email{Santiago.Matalonga@uws.ac.uk}

\renewcommand{\shortauthors}{de França, et al.}

\begin{abstract}
\emph{Context}: As the adoption of continuous delivery practices increases in software organizations, different scenarios struggle to make it scale for their products in a long-term evolution. This study looks at the concrete software architecture as a relevant factor for successfully achieving continuous delivery goals.
\emph{Objective}: This study aims to understand how the design of software architectures impacts on the continuous deployment of their software product.
\emph{Method}: We conducted a systematic literature review to identify proper evidence regarding the research objective. We exploit two search strategies to identify relevant sources. We analyzed the selected sources adopting a synthesis and analysis approach based on Grounded Theory.
\emph{Results}: Through a systematic literature review, we selected 14 primary sources describing elements of concrete architectures that support continuous deployment. Through our analysis process, we developed a theory that explains the phenomenon of Architecting for Continuous Deployment. The theory describes three other phenomena that provide support for Architecting for Continuous Deployment: Supporting Operations, Continuous Evolution, and Improving Deployability. Furthermore, the theory is composed of the following elements: contexts, actions and interactions (strategies, practices, techniques, design patterns), quality attributes, principles, and effects. We instantiated these elements and identified their interrelationships. The theory is supported by providing bi-directional traceability from the selected sources to the elements and vice-versa.  
\emph{Conclusions}: Developing adequate architecture plays a crucial role in enabling continuous delivery. Supporting operations becomes vital to increase the deployability and monitorability of software architecture. These two outcomes require that developers accept responsibility for maintaining the operations. The continuous evolution of the architecture is essential, but it must consider balanced management of technical debt. Finally, improving deployability requires attention to the test strategy and how it affects downtime to enable efficient pipelines.

\end{abstract}



\maketitle

\section{Introduction}
Organizations developing software-intensive solutions that involve innovation or meaningful time-to-market constraints must continuously adapt to untimely changes so they can achieve business goals in the face of a dynamic and competitive market. In this scenario, organizations need to adopt software practices enabling the capacity to deliver new features that can be readily available to end-users. Continuous Software Engineering (CSE) is the umbrella term referring to this set of practices \cite{Bosch2014-ag} \cite{Fitzgerald2017-rg}. 

In principle, several organizations report the benefits of adopting the CSE paradigm. These reported benefits include reduced cycle time, increased reliability productivity, and efficiency \cite{Leppanen2015-qc}. However, the enumerated benefits are not pervasive as others also report challenges when adopting CSE practices, such as increased effort on testing \cite{Leppanen2015-qc, Mantyla2015-fa, Claps2015-lj, Rodriguez2017-wg}, or increased software failure rate \cite{Claps2015-lj,Marschall2007-yx}. Furthermore, we argue that most research has dealt with the interplay between operations and development \cite{Virmani2015-jn,Zhu2016-hu} or has focused on the development tools and deployment pipelines that are needed to support CSE \cite{Shahin2017-yi}. A few of these works have highlighted the importance the concrete architecture has on enabling continuous deployment. For instance, \cite{Rodriguez2017-wg} mentions that flexible product design is a factor for continuous delivery. Similarly, a result of \cite{Laukkanen2017-li}, suggests that some architectures might not be suitable for Continuous Delivery or Deployment (CD). We claim that not enough attention has been paid on designing and developing a software architecture that can support the continuous delivery of its software product. 

This research aims at building a theoretical model of the issues relating to software architecture (including its evolution, decay, and technical debt) and its capacity to support the continuous delivery of the software product. To achieve this, we carried out a systematic literature review (SLR) \cite{Kitchenham2007-dg}. The SLR guided the identification of the available evidence. We adopted a qualitative analysis process based on Grounded Theory (GT) procedures to support the analysis and synthesis of the scientific evidence \cite{Glaser2017-bg}. As a secondary study, we did not follow a whole GT process. Instead, we used GT procedure to support the qualitative synthesis \cite{Dixon-Woods2005-ik} of the evidence revealed by the SLR. 

We identified 14 primary resources that provided evidence about how concrete software architectures affect the capacity to deliver software frequently. As a result of the qualitative analysis process, we developed a preliminary theory that describes the phenomenon of architecting for continuous delivery. It also includes three supporting phenomena that arise when looking at designing and supporting such an architecture. In short, the theory describes the following four phenomena: (1) Architecting for continuous deployment, this is the core phenomenon of the theory. This phenomenon deals primarily with software design concerns, but it also describes the context, symptoms to ensure the continuous delivery of valuable software; (2) Supporting operations, which deals with the software developers’ capacity  to improve the systems’ monitorability and deployability; (3) Continuous evolution, which deals with increasing the ability to deliver software frequently while avoiding cost overrun. And it strikes a balance between the need to deliver software regularly and the management of technical debt; (4) Improving deployability, which deals with developing efficient pipelines that minimize system downtime. Improving deployability stress the impact that a test strategy has in reducing system downtime.

This paper describes a paradigm and a theory instantiated from it by identifying the elements influencing the aforementioned phenomena. We support their evidence-based nature by providing bi-directional traceability between the elements and their sources. We claim that, while the paradigm and the theory might still offer a partial and preliminary picture of all the concepts associated with architecting for continuous delivery, the results presented in this paper are rich enough to provide explanations, articulate hypotheses, and support further research.

\section{Background}
\label{sec:background}

The software industry has shown an increasing interest and adoption for practices establishing a more continuous flow for the software lifecycle, taking into account the need for change on how software has been conceived, developed, and maintained over the last years. Such interest and adoption are especially true in the context of uncertainty and technological innovation. This new thinking regarding the software lifecycle has been named CSE \cite{Bosch2014-ag,Fitzgerald2017-rg} as it includes practices requiring more constant pacing when performed. Besides, it is rooted in agile practices such as short iterations, which fostered rapid releases and made substantial use of automation, mainly in software build and testing activities. Such automation usually happens in the context of Continuous Integration (CI) practice \cite{Beck2000-pb}. With the perceived agility brought by the use of CI, the automated deployment appears as to be an additional step. This way, after successfully executing tests, it makes the application available, in a functional state, and configured into another environment, staging or production. This extension is called Continuous Delivery or Deployment (CD) \cite{Humble2010-rc}.

In this context, methods and practices supporting software development and operations, along with supporting tools, are proposed and released in the market (by industry or academia) aiming at making the achievement of CSE goals easier, particularly w.r.t. time to deliver software products and services \cite{Bosch2014-ag}. However, several technologies have neither proven their effectiveness or efficiency in providing the intended or claimed benefits nor known limitations associated with their continuous use. Additionally, there is not enough evidence to support decision making in the organizations on the adequacy of such technologies to their contexts. It may  cause the failure of introducing them to improve processes concerning delivery velocity, changeability, and the improvement of product or service quality. Some challenges associated with the introduction of such practices include:
\begin{itemize}
    \item Increase effort on testing activities, including test automation, to keep up with constants changes \cite{Leppanen2015-qc,Mantyla2015-fa,Claps2015-lj,Rodriguez2017-wg,Marschall2007-yx, Olsson2012-mx};
    \item Increase of the software failure rate for not having enough time to plan and execute efficient tests \cite{Marschall2007-yx,Claps2015-lj};
    \item Increase of the architectural decay during continuous software evolution \cite{Riaz2009-yz};
    \item Growth of Technical Debt (TD) due to reduced time for planning the implementation of change requests \cite{Mantyla2015-fa};
    \item The domain in which a company operates (telecommunications, embedded systems, games, and others) may impose restrictions on rapid release practices \cite{Leppanen2015-qc, Rodriguez2017-wg};
    \item Software teams resistance to changes required when adopting rapid release processes \cite{Leppanen2015-qc, Rodriguez2017-wg};
    \item Customers unwillingness to deal with frequent releases \cite{Mantyla2015-fa,Claps2015-lj,Rodriguez2017-wg}.
\end{itemize}
 
Particularly, in this study, we are concerned with issues related to software architecture, including its evolution, decay and TD, on the use of rapid release practices, which may impact directly on the delivery capability. In this sense, Rodríguez \textit{et al.} \cite{Rodriguez2017-wg} identified in a systematic review ten groups of factors contributing to the adoption of continuous deployment, which includes flexible product design and architecture as a success factor. Similarly, Laukkanen \textit{et al.} \cite{Laukkanen2017-li} identified problems, causes and solutions related to system architectural design in the context of a systematic mapping concerning the adoption of CD. These include system modularization, unsuitable architecture, internal dependencies, database schema changes. Although both secondary studies agree on software architecture and design as one important success factor for CD, they do not provide further explanation on how these concepts are articulated to improve chances of succeeding in CD initiatives.

Delivery and deployment capability are represented by the minimum frequencies (time) at which software teams can deliver or deploy software artifacts \cite{Makinen2016-vq}. In this sense, two frequency metrics can be used: (1) the actual releasable software cycle, which represents the cycle a development team takes to produce an artifact that could be released, but factors out of control of the team prevent such release; and (2) the actual release cycle means how often the software artifact is actually released. We are particularly interested in the first given perspective and how the software architecture contributes to increase such capability.

Regarding primary studies, Bosch and Eklund \cite{Bosch2012-lk} discuss architectural implications for continuous experimentation. The authors explore concepts related to continuous evolution and experimentation of embedded systems through a case study in the automotive industry. In different domains, Bellomo and colleagues \cite{Bellomo2013-ey} investigated, through an approach based on GT, factors enabling rapid releases. The results show the combination of architecture-related and regular agile practices to balance development speed and product stability.

Usually, the association of software architecture and CSE (more specifically DevOps) ends up with the recommendation of adopting Microservices or Serverless architecture styles. However, we understand these technologies are no silver bullets. After conducting an empirical study, Shahin \textit{et al.} \cite{Shahin2019-bb} proposed a conceptual framework to support the process of architecting for CD. Additionally, they also characterize the principle of “small and independent deployment units”, which is a more abstract expression of the previously mentioned architecture styles. It represents a concrete direction for the relationship between software architecture and CSE, but we understand further investigation is required, including a broader view of this phenomenon.

\section{Research Method}
\label{sec:method}

The main goal of this study is to capture and synthesize evidence from the literature targeting the development of a theory to describe how delivery capability is affected by elements of the software architecture in the context of CD practices. The goal does not necessarily aim at achieving a general theory on this topic, as the achievement of such a goal is dependent on the depth of the existing evidence. Rather, we need a first attempt to synthesize such evidence and to direct efforts on missing empirical results. Therefore, to achieve the goal of this research, we have commissioned a Systematic Literature Review \cite{Kitchenham2007-dg}. An SLR synthesizing scientific evidence from the literature can drive the building of a theory as stated in the research goal. A systematic mapping study (SMS) would not be enough to achieve this goal, as SMS are structured towards organising a research area, while our goal requires in-depth synthesis and analysis. We purposefully exclude gray literature since it is difficult to obtain systematic and unbiased evidence from it. Furthermore, as an inclusion criteria (see Section 3.3), we convey how we are interested only in empirical studies, preferably occurring in real settings. 

To analyse and synthesize the scientific evidence, we followed a qualitative analysis process based on Grounded Theory \cite{Glaser2017-bg}. The suggestion and use of GT as a method for qualitative synthesis of SLR outcomes is recognized outside Software Engineering \cite{Dixon-Woods2005-ik,Finfgeld-Connett2014-fd}. Besides, it is reinforced in the SE \cite{Dyba2007-pv} and Information Systems \cite{Wolfswinkel2013-fn} literature. Several secondary studies in SE have applied GT as a synthesis method \cite{Aleti2013-mh,Garousi2016-nj}. Differently from these works, which focused mainly on categorization, our synthesis also follows a theoretical structure, formalized as an adaptation of the paradigm proposed in \cite{Straus1990-oz}. 
We have made available the research protocol in \cite{De_Franca2020-uy}.

\subsection{Research Questions}

The main research question for this study is ''Which architectural characteristics contribute to the delivery capability?'' Additionally, we breakdown this idea into four secondary questions:

\begin{itemize}
    \item [RQ1] What are the variables concerning the concrete architecture influencing the delivery capability?
    \item [RQ2] How does the quality of the concrete architecture impact the delivery capability?
    \item [RQ3] How do these variables relate to each other to contribute to the delivery capability?
    \item [RQ4] How does the effect of the identified variables evolve over time?
\end{itemize}

\subsection{Search Process and Strategy}

We performed an automated search in the main used libraries in Software Engineering: Scopus, IEEEXplore, and Web of Science\footnote{We decided not to use ACM Digital Library, due to its well-documented reproducibility issues \cite{Giustini2013-rw}.}. Also, we complemented this search with a Snowballing process \cite{Wohlin2014-bf}. Following the results of an \emph{ad-hoc} literature review (summarized in Section \ref{sec:background}), we calibrated the candidate search string using three references as control papers \cite{Bosch2012-lk,Bellomo2013-ey,Shahin2019-bb}. The resulting string was:

\textit{
("continuous* evol*" OR "continuous architecting" OR "continuous software evolution" OR "continuous delivery" OR "continuous deployment" OR "continuous integration" OR "continuous software engineering" OR "rapid releases" OR "rapid fielding") AND ("software architecture" OR "software design" OR "architectural design")
}

Regarding the snowballing process, we followed the guidelines in \cite{Wohlin2014-bf}, applying both backward and forward snowballing. The supporting tools are the same search engines from the automated search. Inclusion/Exclusion criteria for sources identified through Snowballing are also the same as for the automated search strategy (see Section 3.3). Seeds for the snowballing process are the output of the automated search process (after the data extraction phase). Before initiating the snowballing process, the seeds set should be checked against the criteria detailed in the snowballing guidelines (diversity of the community of practice, the size of the seed set, diversity of authors and publishers).

\subsection{Inclusion and Exclusion Criteria}

Inclusion (I) and exclusion (E) criteria follow the research interest on architectural practices supporting CD. Thus, we defined the following criteria:
\begin{itemize}
    \item [I1] Discuss continuous delivery.
    \item [I2] Discuss the software system’s architecture within the CD in terms of its variables.
    \item [I3] Present an empirical/experimental study or experience report with observations in the context of CD.
    \item [E1] Gray literature.
    \item [E2] Papers not written in English.
    \item [E3] The same study reported twice, only the most complete report will be considered.
    \item [E4] Paper presenting no empirical/experimental evidence (only proposal or position papers).
    \item [E5] Papers discussing the architecture of the deployment pipeline, including its testing harness.
\end{itemize}

\subsection{Selection Procedure}

We followed the same selection procedure for both the automated and the snowballing search. All authors participated as reviewers equally. To calibrate our interpretation of the criteria, we adopted an iterative process for the search string development. With the resulting search string, we selected 10 sources to be reviewed and the application of I/E criteria by each researcher discussed.
The resulting set of papers was divided into thirds. Each researcher reviewed two-thirds of the papers. The selection process will be guided by Table \ref{tab:selproc} (adapted from \cite{Petersen2015-if}). Papers classified as F will be excluded. Papers classified as C and D will be read in full to justify inclusion/exclusion.

\begin{table}[]
  \caption{Selection Procedure}
  \label{tab:selproc}
\begin{tabular}{|c|l|l|l|l|}
\hline
\multicolumn{2}{|c|}{\multirow{2}{*}{}} & \multicolumn{3}{c|}{Reviewer Y}                                                              \\ \cline{3-5} 
\multicolumn{2}{|c|}{}                  & \multicolumn{1}{c|}{Include} & \multicolumn{1}{c|}{Uncertain} & \multicolumn{1}{c|}{Exclude} \\ \hline
\multirow{3}{*}{Reviewer X} & Include   & A                            & B                              & C                            \\ \cline{2-5} 
                            & Uncertain & B                            & C                              & D                            \\ \cline{2-5} 
                            & Exclude   & C                            & D                              & F                            \\ \hline
\end{tabular}
\end{table}

\subsection{Data Extraction}

Table \ref{tab:extractionform} presents the data extraction form developed to answer our research questions. 

\begin{table}
  \caption{Information Extraction Form}
  \label{tab:extractionform}
  \begin{tabular}{p{0.30\linewidth}p{0.50\linewidth}p{0.10\linewidth}}
    \toprule
    Field & Description & Related RQ\\ 
    \midrule
    Bibliographic information & Title, authors, abstract, publication venue and date. & N/A \\
    Variables and quality metrics influencing delivery capability & Architecture-related factors, moderators, and mediators impacting on delivery capacity, as well as their description and how they are measured. & RQ1 \\
    Direction of the influence & Positive or Negative. Explain & RQ2-3 \\
    Measurement/Metric for delivery capability & Metrics defining delivery capability & RQ2 \\
    Evolution perspective & Does the paper discuss variables in the evolution perspective (over time)? & RQ4 \\
    Type of study & Controlled experiment, case study, action-research, other observational studies. & N/A \\
    Contextual information & Characteristics from organizations, products, teams, and projects. & N/A \\
    Architectural elements & Passage detailing the architectural elements or issues & RQ1 \\
    Concrete architecture & Evidence of implementation of the architectural style into a concrete architecture & RQ1 \\
    \bottomrule
  \end{tabular}
\end{table}

\subsection{Quality Appraisal Criteria}

We derived our quality criteria (Table \ref{tab:quality}) based on \cite{Runeson2009-sg} and \cite{Kitchenham2012-ak}, so that we included a more qualitative set of criteria for evaluating quality and a more experimental set, respectively.

\begin{table}
  \caption{Quality appraisal criteria}
  \label{tab:quality}
  \begin{tabular}{p{0.75\linewidth}}
    \toprule
    \textbf{Quality Criterion}\\
    \midrule
    \textit{Category: questions on aims}\\ 
    Are the objectives, research questions, and hypotheses (if applicable) clear and relevant?\\
    Is the suitability of the case to address the research questions clearly motivated?\\    
    \midrule
    \textit{Category: questions on design, data collection, and data analysis}\\
    Do the authors describe the cases (samples or experimental units)?\\
    Do the authors describe the design of the study?\\
    Do the authors describe the data collection procedures and deﬁne the measures?\\
    Do the authors deﬁne the (quantitative/qualitative) data analysis procedures?\\
    Is a clear chain of evidence established from observations to conclusions?\\
    Are threats to validity analyses conducted in a systematic way and are countermeasures taken to reduce threats?\\
    \midrule
    \textit{Category: questions on study outcome}\\
    Do the authors state the ﬁndings clearly?\\
    Is there evidence that the study can be used by other researchers/practitioners?\\
    \bottomrule
  \end{tabular}
\end{table}

\section{Conducting the Review}
\label{sec:conducting}

\begin{figure}[ht]
  \centering
  \includegraphics[width=\linewidth]{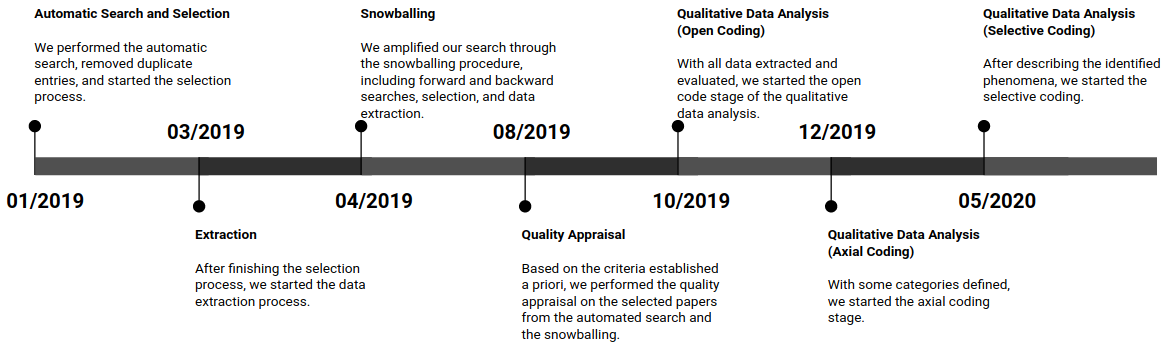}
  \caption{SLR Process Timeline.}
  \label{fig:timeline}
\end{figure}

This section presents general issues with SLR process’ execution, as mentioned, the full protocol is available at \cite{De_Franca2020-uy}. Figure \ref{fig:timeline} presents the timeline of the research process. As shown, some activities were performed several times on different sets of sources. For illustration purposes, this section gives only representative samples to convey decisions taken during the research process.

\subsection{Automatic Search Process}

The automatic search process was carried out between January and February 2019. Tables \ref{tab:papersengine} summarizes the results.

\begin{table}
  \caption{Identified papers by search engine}
  \label{tab:papersengine}
  \begin{tabular}{p{0.25\linewidth}p{0.06\linewidth}}
    \toprule
    Engine & Papers\\ 
    \midrule
    Scopus&444\\
    IEEE&79\\
    Web of Science&46\\
    Total&569\\
    Removing Duplicates&476\\
    Ready for selection (removing non-papers entries)&468\\
    \bottomrule
  \end{tabular}
\end{table}


From the 468 papers, we selected 25 for the full reading and extraction. As shown in  Table \ref{tab:selproc}, all researchers participated in the voting process. We used JabRef reference manager as a tool to support this activity during the automated search process.

From these 25 sources, we excluded 13 papers after the full reading. Mostly, these exclusions were motivated by (1) the lack of presentation or discussion of architectural elements, for instance, see \cite{Balalaie2016-yj}; and/or (2) the lack of an empirical/experimental study providing evidence, for instance, see \cite{Chen2015-jz}. 

Data extraction was performed verbatim for all twelve papers remaining. We used templates in word processing software to instantiate the data extraction form previously presented (Table \ref{tab:extractionform}). To assure consistency, the research team reviewed all extracted data.

The aforementioned process, including quality checks, was followed for both the sources identified through the automatic search and the snowballing search process.

\subsection{Snowballing}

The 12 sources included in the automated search were used as seeds to the snowballing process. Following the guidelines by \cite{Wohlin2014-bf}, we reviewed the suitability of these 12 sources to act as seeds.  Out of the 12 seeds, three sets of authors repeat (Bellomo, Chen, and Shahin and Babar). Nonetheless, we decided to keep all papers from the automatic search method to use as seeds because:
\begin{itemize}
    \item The two papers by Bellomo look at different aspects.
    \item Though papers from Chen and Shahin and Babar might describe the same research, the difference in publication year might result in losing forward citations if only the newest ones are selected (y. 2018).
\end{itemize}

We used both forward and backward snowballing in \cite{Wohlin2014-bf}. For each source, two researchers applied the selection criteria using the paper titles in their references section. Besides, we noted and discussed discrepancies. Likewise, we searched for papers citing each source in the aforementioned search engines (forward snowballing) and applied the inclusion/exclusion criteria.

From these references and citations, we selected 21 candidates from the backward snowballing and 18 from the forward. Finally, we reviewed these 35 candidates in full reading and, based on the selection criteria, and two new references were included in the final set for analysis, resulting in fourteen papers in total (see Table 8).

\section{Findings}
This section describes the outcomes of the SLR process, conveying factual information about our findings. Synthesis and interpretation is left for later sections.

\subsection{Overview}
\begin{figure}[h]
  \centering
  \includegraphics[width=0.7\linewidth]{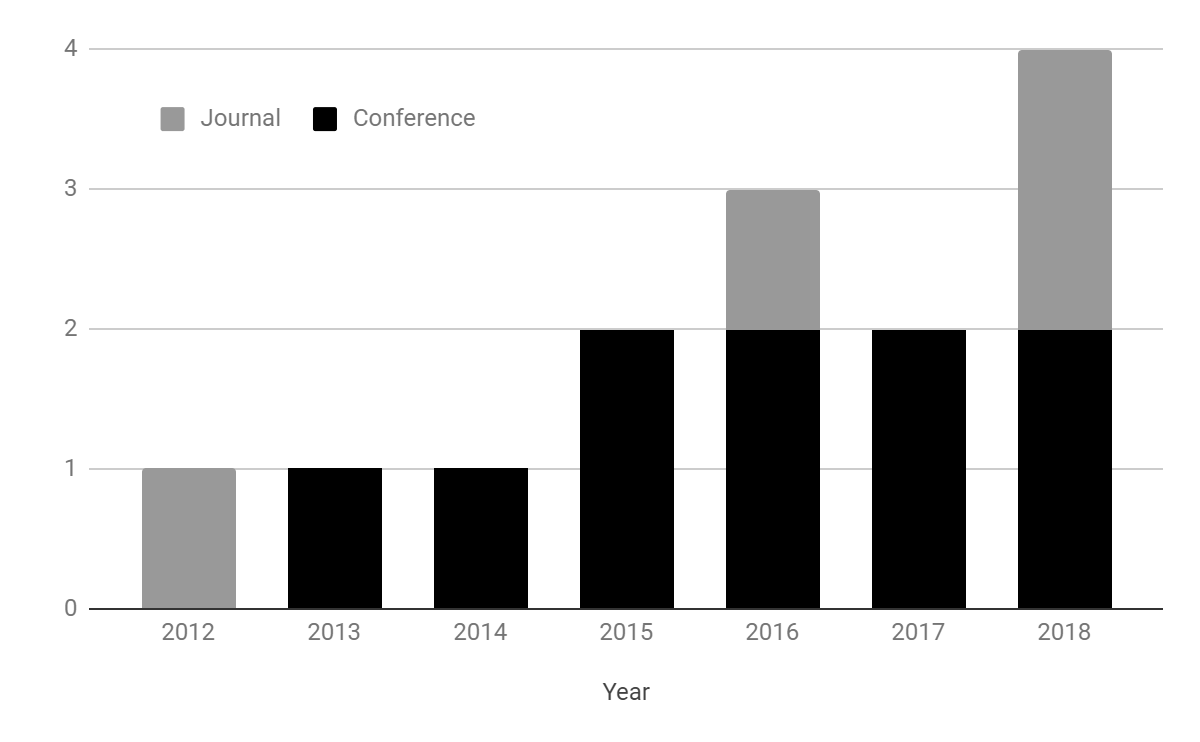}
  \caption{Publication year distribution}
  \label{fig:distribution}
\end{figure}

The studies’ publication period ranges from 2012 to 2018. The median for the cumulative publication frequency data is 2016. Figure \ref{fig:distribution} shows the overall distribution over the years in an  ascending trend. Publication venues listed in Table \ref{tab:venues} are diverse with almost one paper per venue, except for the Conference on Software Architecture, with three publications, which is the main subject of this paper. It is probably due to the small number of publications, but it calls attention to the variety of topics in which software architecture and delivery capability have been addressed, including the Internet of Things, Dependable Systems, and Software Process.

\begin{table}
  \caption{Publications venues and frequency}
  \label{tab:venues}
  \begin{tabular}{p{0.75\linewidth} c}
    \toprule
    Publication Venue&Frequency\\ 
    \midrule
    Working Conference on Software Architecture (WICSA) and International Conference on Software Architecture (ICSA)&3\\
    International Conference on Software Engineering (ICSE)&1\\
    International Conference on Dependable Systems and Networks (DSN)&1\\
    Hawaii International Conference on System Sciences (HICSS)&1\\
    International Conference on Product-Focused Software Process Improvement (PROFES)&1\\
    International Conference on Internet of things, Data and Cloud Computing (ICC)&1\\
    International Symposium on Empirical Software Engineering and Measurement (ESEM)&1\\
    Colombian Conference on Computing (CCC)&1\\
    \textbf{Total of conference papers}&10\\
    Empirical Software Engineering (EMSE)&1\\
    Information \& Software Technology (IST)&1\\
    IEEE Software&1\\
    CrossTalk: The Journal of Defense Software Engineering&1\\
    \textbf{Total of journal papers}&4\\
    \textbf{Total}&14\\
    \bottomrule
  \end{tabular}
\end{table}



\subsection{Identified Studies}

Bachmann et al. \cite{Bachmann2012-ik} discuss lessons learned from software projects concerned with how architectural tactics can sustain rapid and agile software development. The authors point out three distinct tactics that can provide the degree of architectural stability required to support the next iterations of development: (i) align feature-based development and system decomposition, (ii) create an architectural runway and (iii) use matrix teams. From these tactics, the first two concern software architecture. Feature-based development and system decomposition alignment are essential to ``\emph{create a platform containing commonly used services and development environments either as frameworks or platform plug-ins}'' to develop new features upon the common platform. In the architectural runway, the \emph{``agile teams build a runway of infrastructure sufficient to support the development of features in the near future.''} 

Bellomo et al. \cite{Bellomo2013-ey} present a comprehensive investigation regarding the factors that enable or inhibit ``rapid fielding.'' 
It seems to be a concept borrowed from the military jargon, which resembles our understanding of (rapid) delivery capability. The authors conducted an observational study based on interviews, focusing on projects adopting an iterative process to investigate the ``rapid fielding'' influencing factors. Although the research is comprehensive regarding the factors affecting rapid fielding, some aspects are specifically related to software architecture. Interviewees from two organizations said they are starting to keep a list of design decisions along with an incremental architectural change plan that could reduce the likelihood of a big-bang response, which could affect the delivery capability. Also, it is possible to identify a concern with making architectural problems visible to the business area.

Bellomo et al. \cite{Bellomo2014-ai} present a similar research goal and methodology compared to their previous work. However, it focuses on architecture decisions for achieving deployability goals. 
Interviewees from three software projects with different deployability goals engage the study. Four general deployability goals, also called continuous delivery goals, are identified as enabling: (i) build automation and continuous integration, (ii) test automation, (iii) deployment and robust operations, and (iv) synchronized and flexible environments. From these goals, the authors gathered a set of architectural decisions. Respectively to the four general deployability goals, some examples of deployability tactics include (i) maintenance of existing interfaces, (ii) encapsulation, (iii) to increase computational efficiency, and (iv) virtualization.

Chen \cite{Chen2015-jz} provides lessons in migrating their custom software applications to CD. Those of most interest are related to what architecture implies to CD. Four requirements affecting the software systems' architecture are identified and discussed. Deployability is the first mentioned because software applications are ``deployed to several testing environments multiple times a day and deployed to the production environment once or twice a week.'' Security is mentioned since ``the security vulnerabilities of the application during the start-up time become more important, as attackers (hackers) get more opportunities to attack the software during its start-up time.'' Another important requirement is loggability given that the architecture must be prepared to log “sufficient information for diagnosis and troubleshooting when issues arise” and, at the same time, make logs ``concise enough to not log anything that does not justify its cost.” Lastly, modifiability is mentioned as an essential requirement “to allow constant incremental adding of small new features.''

Villamizar et al. \cite{Villamizar2015-hs} present lessons learned when migrating a monolith application to microservice architecture in a cloud environment. Because of the particular context of cloud computing, more precisely using the Amazon Web Services, the authors focus on more technical aspects of the application on a cloud service such as configuration, performance, and economic aspects. Regarding delivery capability, the authors state that ``the use of continuous delivery strategies in microservices can be a time-consuming task, due to repetitive and manual tasks must be executed in each deployment; the use of automation tools are mandatory to save time and gain agility.'' These observations should be interpreted taking cloud computing into account. Nevertheless, automation appears as an important factor affecting delivery capability.

Balalaie et al. \cite{Balalaie2016-yj} also investigate the microservice architecture in the context of cloud computing. They report the experiences of PegahTech in migrating Backfactory, a commercial mobile backend as a service, to microservices in the context of DevOps. The paper does not focus on specific delivery capability aspects related to the architecture, even though it is possible to identify several elements in the study. For instance, the migration was driven by the need for decentralized data governance and automated deployment. The former, because shared databases affect deployability. And the latter due to a more significant number of deployable units (i.e., microservices). Furthermore, several architectural patterns are used in the migration because of the microservice architecture. These include bounded context, circuit breaker, load balancing, and service discovery, which also affect deployability. 

Chen et al. \cite{Chen2018-ih} conduct a participatory case study to understand and adapt agile analytics practices to the context of big data systems. The main goal was to allow data scientists and other stakeholders to increase value discovery amount and frequency. The authors identified software architecture methods as a key enabler for addressing this issue as they can address both business and technical aspects. The method Architecture-centric Agile Big data Analytics (AABA) was developed collaboratively over 10 cases (i.e., big data projects) to provide a basis for value discovery, planning and estimating cost and schedule, supporting experimentation, and enabling continuous delivery of value. The method includes an iterative design method for big data systems based on well-known design primitives (patterns, tactics, references architectures, and technologies). 

The study in \cite{Shahin2016-pp} is one of the closest related to the goal of our synthesis study. They assert that ``a critical dimension of CD is to explore and understand the role of software architecture in transition to CD practice, i.e., how software should be (re-) architected to enable and support CD principles (e.g., frequent and reliable deployment).'' The study interviewed 16 participants from different organizations. The interviews focused on identifying architectural principles necessary for the CD practice, which are enumerated here: (i) small and independent deployment units (e.g., microservices), (ii) not too much focus on reusability (to reduce dependencies), (iii) aggregating logs (using external tools to aggregate and abstract the log data over time), (iv) isolating changes (using practices such as bounded context from domain driven design and feature toggles), and (v) testability inside the architecture (to support automation).

Lehmann and Sandnes \cite{Lehmann2017-wf} propose a framework designed to support architects in selecting a strategy and technology stack for implementing continuous delivery. Their framework on microservice-based software systems, and its organization is based on two interviews and existing literature. Based on a set of criteria considered relevant to the development and delivery, they enumerate five criteria: (i) testability, (ii) deployment abstraction (related to the level of automation), (iii) environment parity (development, testing, and production), (iv) time to deploy (in minutes), and (v) availability adequacy (due to deployment or resource scaling). Except for time to deploy, all criteria have a three-level scale to assess the continuous delivery strategy. For instance, environment parity has different levels according to the number of manual steps and configurations necessary to run the system on each environment. The three levels are disparate (in which parity is ensured manually on development, testing, and production machines), distinguishable (where some differences do exist but can be easily mitigated), and equal (in which the same outcomes are expected from the software in any environment).

Martenssom et al. \cite{Martensson2017-up} investigate impediments for implementing continuous integration. They interviewed 20 experienced software developers asking them to describe the impediments and later showing them a list of impediments found in technical literature. Regarding software architecture, the main factors affecting continuous integration practices are related to ``system thinking,'' which includes ``modular and loosely coupled architecture.'' As mentioned in several studies, the respondents commented that a modular architecture avoids problems where many teams want to make changes in the same component, favoring to break down the work into small chunks and working in parallel. 

The study in \cite{Chen2018-ih} is an extension of \cite{Chen2015-jz} discussing challenges that emerged after adopting a microservice architecture. These new challenges are associated with the increased number of services, evolving contracts among services, technology diversity, and testing. Contracts among services, for instance, has a large impact on delivery capability. The robustness principle, related to the contracts among services, defines that services must be conservative in what they send and liberal in what they accept so that they can evolve without breaking compatibility. Also avoiding breaking compatibility, the expand and contract principle adds new interfaces instead of modifying them to allow consumers to migrate the new interfaces gradually. The author concludes with considerations regarding the added costs due to these new complexities and challenges. According to their experience, building the CD platform took a team of eight people four years. And for that reason, ``the microservices approach is for handling complex systems that require high speed changes.''
	
Ivanov and Smolander \cite{Ivanov2018-lq} explore the influence of serverless architectures on DevOps practices, focusing on CI, CD, and monitoring. In this study, a build pipeline was designed and implemented using the Design Science Research methodology in sprints of two weeks. In the sprints, the study participants discussed in a workshop the relevant technical considerations for designing and implementing the pipeline. The findings of the discussions in workshops showed that 18 out of 27 automation practices are affected by the serverless architecture. It includes testing and QA (e.g., mockups and proxies), monitoring (e.g., log aggregation), and build process (e.g., build artifacts managed by purpose-built tools) practices. 

Schermann et al. \cite{Schermann2018-og} focus on empirically characterizing the state of the practice of continuous experimentation practices. Even though continuous experimentation is not the theme of this synthesis, it has a large intersection with CD. They adopt a research methodology similar to the other studies using interviews and a survey. One of the technical practices identified in the study is the necessity of automation for CI and CD. Regarding software architecture and its implications for CD (and by extension, continuous experimentation), the findings show the importance of a loosely coupled architecture usually materialized in the form of microservices. Also, the study reveals that feature toggles are important mechanisms to circumvent architectural limitations.   

Shahin et al. \cite{Shahin2019-bb} conduct interviews and a survey to address a similar goal of this research, which is how software systems should be architected for continuous delivery and deployment. One important concern of their research is acknowledging that monolithic architecture is predominant in software industries. By proposing a conceptual framework to support (re-)architecting a system for CD, they indicate architectural attributes that should be augmented/improved if organizations want to achieve CD with, for instance, their monoliths. The main findings of the study are: (i) ``small and independent deployment units'' is an alternative to monolithic systems and represent a foundation to design CD-driven architectures, (ii) monoliths and CD are not intrinsically oxymoronic, and its challenges can be mitigated (e.g., reducing test run times by improved test quality), (iii) identification of first-class quality attributes in the CD context (i.e., deployability, modifiability, testability, monitorability, loggability, and resilience), and (iv) CD requires operations-friendly architectures. 

\section{Qualitative Analysis and Synthesis}

As data collection, we performed the SLR process described in Sections \ref{sec:method} and \ref{sec:conducting}. Thus, primary data sources are the fourteen original papers, from which we extracted relevant qualitative information according to the data extraction forms. Even though 14 papers might not seem like a large sample, the amount of qualitative evidence uncovered by the data extraction process needs a systematic method for its synthesis. 

Since we target to develop an initial theory on architecting for CD, we considered different perspectives on which type of theory we could achieve. Mainly considering the type of empirical studies obtained from the SLR, we adopted a synthesis and analysis approach based on GT \cite{Straus1990-oz}. Grounded Theory has been extensively exploited for data synthesis in software engineering studies \cite{Carver2007-ce}\cite{Cruzes2011-zn}, though its application has some criticism \cite{Stol2016-qn}. In this study, we take cue from \cite{Stol2016-qn} and provide evidence of our application of GT procedures.

We aim to develop an initial theory on architecting for CD as a result of the synthesis. Since there are several theory traditions, it is convenient to explicitly state which one we are aligned to in order to avoid misinterpretations. Three theory traditions are pervasively discussed in the technical literature, namely \cite{Gioia1990-en,Lynham2002-kr,Gay2011-kj}: (1) hypothetico-deductive (alternately referred to by theorists as nomothetic, positivism, or empirical-analytical); (2) inductive-synthesis (alternately referred to as idiographic, grounded theory, constructivism, or interpretive theory); and (3) critical theory (alternately referred to as radical, neo-Marxist, or social justice theory). Given the type of data collected in the primary studies and the synthesis strategy used in this research, we are constrained to adopt the inductive-synthesis view. Its primary goal is to depict what is occurring in a particular situation, by clarifying meanings and interpretations. Furthermore, it should be highlighted what we mean by an initial theory. Carroll and Swatman \cite{Carroll2000-ng} indicate theories have several development levels, from minor working relationships, which are concrete and based directly on observations, to all-embracing theories that seek to explain. Therefore, due to the diverse and relatively small set of 14 primary studies, we understand that the theory developed in this research seeks to describe the entities and working relationships, which are all grounded on the evidence uncovered by the data extraction.

Initially, we familiarized ourselves with the data and performed Open Coding, which is the process of fracturing the data to identify categories, properties, and local dimensions \cite{Straus1990-oz}. Then, we assigned each researcher a set of fields in the data extraction form to keep the consistency of interpretation. That researcher would perform the open coding of the 14 data sources considering just the chunks of text in the fields assigned to him. Work progressed in timeboxed intervals of two weeks, where the research team discussed uncovered codes. This process was supported using QDA Miner Lite\footnote{https://provalisresearch.com/products/qualitative-data-analysis-software/freeware/ },and it led us to the identification of four significant phenomena, explained later in the next section. 

Based on the major categories, we started the Axial Coding, which is the process of reassembling/regrouping codes by relating categories to subcategories to explain the phenomena of interest \cite{Straus1990-oz}. Mainly, axial coding was carried out in plenary meetings by the research team, where the reviewers discussed the axial codes and the pertinence of their associated open codes. The first round of the axial coding resulted in a hierarchy of codes relying heavily on a classification that did not have a good fit in describing and explaining the phenomena. So, we decided to develop a paradigm, similar to the one in \cite{Straus1990-oz}, to organize subcategories and clarify the explanations. This paradigm provides an abstract theoretical structure that allows to describe the synthesized knowledge. Thus, each identified concept plays the role of a construct in the paradigm, and these concepts are related according to associated propositions. The definition of such a paradigm allows us to systematically explain each phenomenon, as well as to check for consistency and missing constructs in the developed theory.  As a tool to document this process, we used Kumu.io. For each category representing a phenomenon, we created memos and updated as it evolved.

Both open and axial codes were discussed and thoroughly reviewed in the plenary meetings. We repeated these activities until the analysis produced no new codes or no re-organization of the axial codes. We do not claim to achieve theoretical saturation as we did not perform additional data collection. However, from the data sources, we could not identify new codes emerging.

Finally, with each phenomenon explained, we searched for the core phenomena (category) and possible relationships among them in the Selective Coding, which is the process of systematically relating the categories and validating their relationships \cite{Straus1990-oz}. We identified the core phenomena in the plenary meetings, and then researchers searched for links explaining the overall theory grounded on data. In this step, we adopted both Kumu.io (to analyze categories) and QDA Miner Lite (to search for groundedness). 

As an SLR, the research goal is to synthesize evidence. In this sense, the developed theory is limited to the explanation capacity and depth of the identified primary studies . In other words, there is no possibility of emerging a theory contradicting the literature or presenting something completely unprecedented. The theory is solely grounded on the sources of empirical evidence. Once all theoretical structures emerge from the empirical evidence, we understand these elements resonate with the experience of software architects (participants from primary studies), contributing to theoretical validity.

\section{A Theory of Architecting for Continuous Delivery}

\begin{figure}[h]
  \centering
  \includegraphics[width=1.0\linewidth]{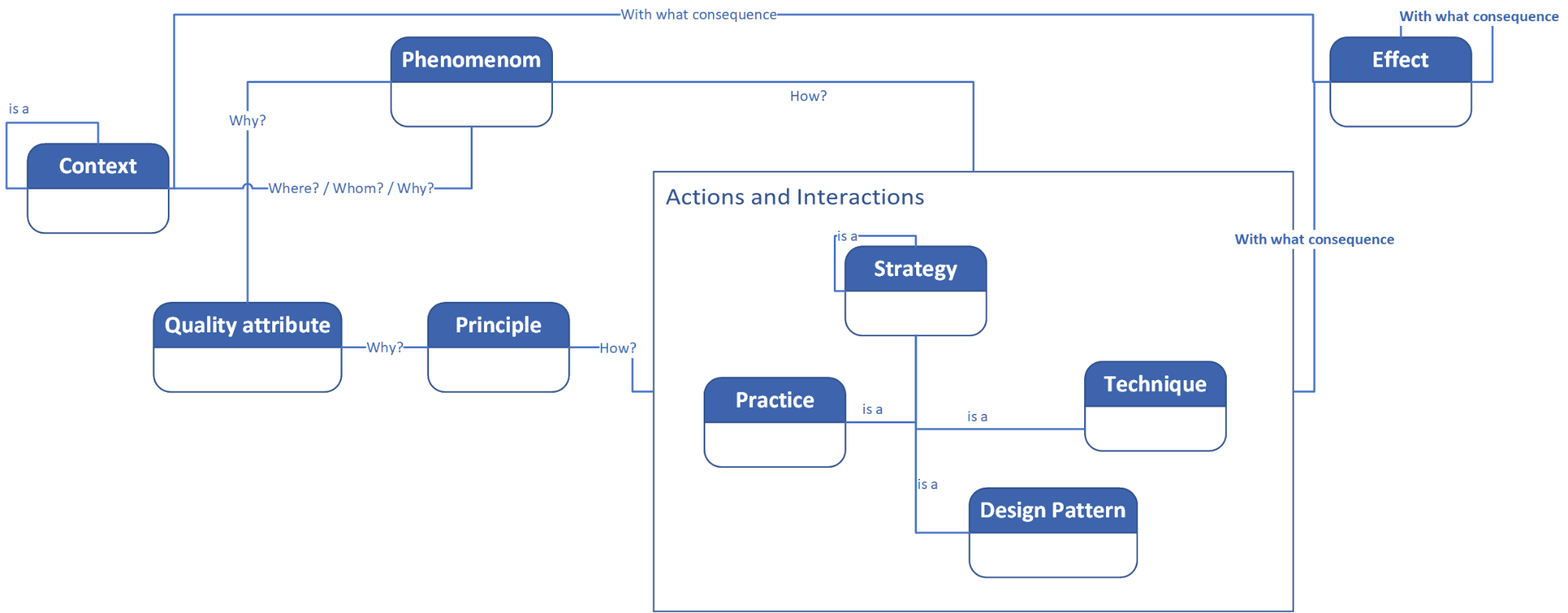}
  \caption{Paradigm developed for the grounded theory}
  \label{fig:paradigm}
\end{figure}

This section presents the grounded theory of Architecting for Continuous Delivery. Figure \ref{fig:paradigm} presents the paradigm which evolved through our application of grounded theory.

This paradigm contains the following element types:
\begin{itemize}
\item Phenomenon: Represent the central empirical situation that emerged in the analysis. There are only four phenomenons identified in our theory, and we will explain them in this section.
\item Context: These elements form the setting to give meaning to other elements. For example, ``type of system'' or ``cost overrun''.
\item Actions and interactions: These are not present in the instance of the paradigm, but it’s an abstraction that we included to simplify the design of the paradigm. There are four element types in this abstraction.
\begin{itemize}
    \item Strategy: A high-level action to achieve an effect. For instance, \textit{``Applying Design Patterns''} is a strategy to enable \textit{``Modularization''} (one of the five phenomenons in our theory).
    \item Practice: Represent routine procedures to achieve a Strategy
    \item Technique: A technique is a determined way of implementing a \textbf{Practice} or \textbf{Strategy}.  Examples of identified techniques are Feature Toggle or Container diagrams.
    \item Design Pattern: Eight design patterns appeared named in our theory; they are all related to \textbf{Strategy} \textit{``Applying design pattern''} with a \textbf{``Is A''} relationship.
\end{itemize}
\item Quality Attribute: A desirable property of the system \footnote{We purposely deviate from the ISO/IEC 25010:2011 \cite{ISO_Standard25010} definition of quality attribute as our understanding of this term in this theory is phenomenological not theoretical.}
\item Principle: Represent agreed-upon beliefs for which there is a shared understanding of the sources. Examples include Conway's Law, the single responsibility principle, and design for testability. 
\item Effect: These are the observed effects of adopting the Strategies, Principles, Practices and Techniques related to Architecting for CD. 
\end{itemize}

The previous element types are connected through edges that have the following semantic:

\begin{itemize}
    \item Why: these relationships try to explain reasons or motivations for the phenomenon to happen. For instance, we interpret \textit{Modifiability} (\textbf{quality attribute}) as a motivation or a target property when performing the \textit{Architecting for CD} \textbf{phenomenon}, so these elements are connected using a \textbf{why} relationship.
    \item With what consequence: these relationships represent what are the implications when a phenomenon occurs. For instance, the \textbf{practice} of \textit{Creating readable logs} achieves improvement on the system \textit{Monitorability} (\textbf{effect}), when \textit{Supporting Operations} (\textbf{phenomenon}).  This way, the \textbf{practice} is connected to the \textbf{effect} using a \textbf{with what consequence} relationship.
    \item How: these connections explain how a phenomenon happens through a set of actions/interactions/interventions. For instance, the \textbf{strategy} of applying \textit{Deployability Tactics} is a way of \textit{Improving Deployability} (\textbf{phenomenon}). Thus, these elements are connected using a \textbf{how} relationship.
    \item Whom: these relationships refer to contextual elements such as actors who are involved in a phenomenon. For instance, in the \textit{Continuous Evolution} (\textbf{phenomenon}) of software systems, \textit{Experienced software developers} are key. Therefore, these elements are connected using a \textbf{whom} relationship.  
    \item Where: these relationships refer to contextual factors such as locations, situations, or conditions describing the phenomenon. For instance, the different \textit{Types of Systems} (\textbf{context}) in which evidence on the \textit{Architecting for CD} \textbf{phenomenon} are found. This way, these elements are connected using a \textbf{where} relationship.
    \item Is A: these relationships describe forms of specializations among elements. For instance, the \textit{Serverless Architecture} (\textbf{strategy}) is a specialization for \textit{Adopting Architectural Styles} (\textbf{strategy}). So, these elements are connected through an \textbf{is a} relationship.
\end{itemize}

In addition to the semantic in theory, a connection can be reinforcing (positive), undermining (negative) or neutral, when we found no evidence to suggest either of the previous classifications.

\subsection{The theory in a big picture}
\begin{figure}[h]
  \centering
  \includegraphics[width=1.0\linewidth]{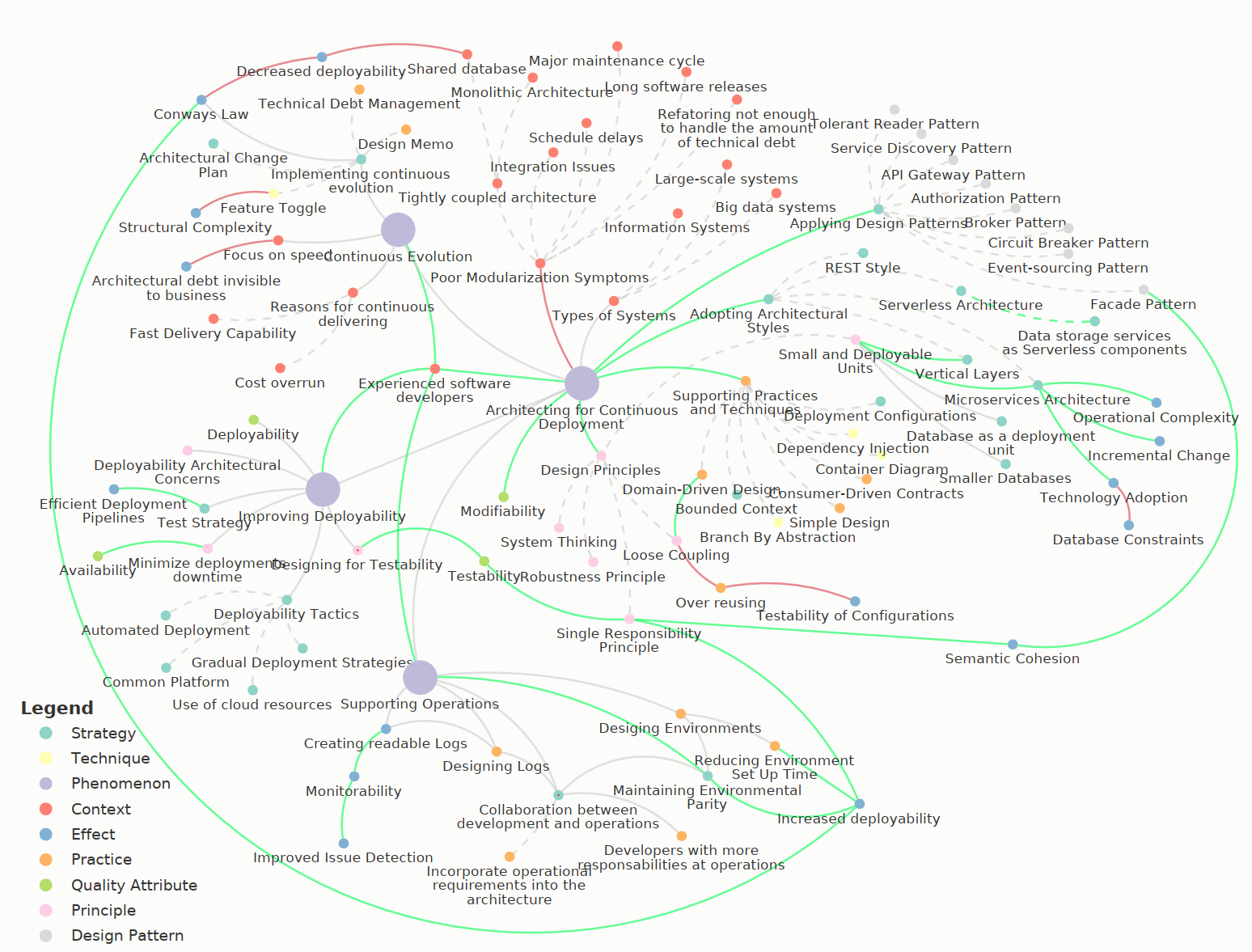}
  \caption{Paradigm developed for the grounded theory}
  \label{fig:BigPicture}
\end{figure}

Figure \ref{fig:BigPicture} shows the grounded theory in an overview\footnote{The diagram is available at: https://kumu.io/brenofranca/architecting-for-cd-selective-coding}. Mainly, it involves four phenomena (larger circles) providing the following explanation: ``\textbf{Architecting for Continuous Deployment}  (core phenomenon) involves software design concerns primarily, but it should also target \textbf{Improving Deployability} (support phenomenon) to ensure the continuous delivery of valuable software. In this sense, \textbf{Supporting Operations} (support phenomenon) is key to reach downstream activities frequently successfully. Also, decisions made for the system architecture impacts its \textbf{Continuous Evolution} (support phenomenon) in the long-term perspective.” The following sections describe each phenomenon in detail.

\subsection{Architecting for Continuous Deployment}
This is the core phenomenon of the developed theory, and the following memo (in the textbox) describes it textually in an overview, emphasizing (in bold) the main involved concepts (open codes). 

\noindent\fbox{%
    \parbox{\textwidth}{%
        In the face of symptoms like\textbf{ integration issues} and \textbf{refactoring not being enough to handle the amount of technical debt}, leading to \textbf{major maintenance cycles, long software releases}, and \textbf{schedule delays}, which are probably caused by the adoption of a \textbf{tightly coupled architecture}, \textbf{Architecting for Continuous Delivery} is a phenomenon occurring in different \textbf{types of systems}. In this context,\textbf{ experienced software developers} adopt different strategies to ensure \textbf{modifiability}, which is a factor for \textbf{delivery capability} by implementing \textbf{design principles}. For that, important strategies take place like \textbf{adopting architectural styles}, such as \textbf{Microservices} and \textbf{Serverless architectures}, \textbf{applying design patterns} on the implementation of these styles, and using \textbf{supporting practices and techniques}, such as \textbf{Domain-Driven Design} and \textbf{Consumer-Driven Contracts}.
    }%
}

From the collected sources, the experiences on architecting for CD are mainly motivated by poor modularization symptoms  \cite{Chen2015-jz}  \cite{Bellomo2013-ey}. Bellomo et al.  \cite{Bellomo2013-ey}  mention major \textbf{maintenance cycles} and \textbf{refactoring not enough to handle the amount of technical debt} as examples of such symptoms. Additionally, poor modularization symptoms also regard \textbf{tightly coupled architectures}, like \textbf{shared databases} and \textbf{monolithic architectures} \cite{Shahin2016-pp}.  Other symptoms include schedule delays and integration issues \cite{Bachmann2012-ik}.

The memo presents \textbf{modifiability} as a factor for \textbf{delivery capability} \cite{Chen2015-jz}. Furthermore, the \textbf{modifiability} quality attribute motivates architecting for CD, which explains why several deployability tactics in \cite{Bellomo2014-ai} are, actually, new interpretations of modifiability tactics. Furthermore, \textbf{design principles} drive the achievement of modifiability through design patterns, architectural styles, and practices.

CD is a practice targeting the delivery of valuable software in frequent and constant pacing. As the product team deploys software releases into production regularly, it is not feasible to make it with larger software modules. This way, an important principle guiding CD is to work with \textbf{small and deployable units} \cite{Shahin2019-bb}. This principle is a step further in the balance of other known principles like the \textbf{single responsibility principle} (SRP) and \textbf{loose coupling} towards a deployment mindset. 

As a consequence of the \textbf{small and deployable units}  principle, we found suggestions of having \textbf{smaller databases} \cite{Shahin2016-pp}. Also, in this same source, we found that \textbf{databases should be treated as deployment units} \cite{Shahin2016-pp}. Another consequence regards the \textbf{over reuse} of components, which can lead to  (1) increased inter-team dependency and (2) losing testability \cite{Shahin2019-bb}.

In addition, Bellomo et al. \cite{Bellomo2014-ai} shows how these principles work together describing that \textbf{SRP improves testability and deployability}. The authors express that when teams achive the modularization of features, \textit{``then lower coupling, and increased cohesion enable deployment and continuous delivery...''}\cite{Bellomo2014-ai}.

Furthermore, we identified a few occurrences of the \textbf{robustness principle} and \textbf{system thinking} as relevant when architecting for CD \cite{Chen2018-ih}.  Martensson et al. mention \textbf{System thinking} as a surprise, since it leads to designing the entire system rather than using a simple design and evolving it by integrating small increments \cite{Martensson2017-up}. 

As mentioned in the memo, this phenomenon has three types of actions/interactions when architecting for CD: adopting architectural styles, applying design patterns, as well as supporting practices and techniques. These actions and interactions are somewhat general for software architecting. However, the specific styles, patterns, practices, and techniques make the difference in the CD context. Architectural styles in the context of CD mainly include \textbf{Microservices} (with \textbf{REST}) \cite{Shahin2016-pp} and \textbf{Serverless} \cite{Ivanov2018-lq}.


Chen \cite{Chen2018-ih} reports an increased \textbf{operational complexity} when adopting the microservices architecture. This consequence is also reported in \cite{Shahin2016-pp} when discussing architectures for CD, explicitly mentioning operations overhead as an example of cause for negative results, culminating in the failure to achieve the potential benefits. On the other side, Chen \cite{Chen2018-ih} also reports that microservice facilitates \textbf{incremental changes}. Another benefit of adopting microservices is also reported by \cite{Schermann2018-og} regarding that it facilitates \textbf{technology adoption}. As a consequence of the former benefit, microservices also reduces database constraints \cite{Chen2018-ih}. Besides microservices, Shahin et al \cite{Shahin2019-bb} identifies the use of the \textbf{vertical layers} style as a means to achieve the principle of \textbf{small and independent deployment units}.

Some design patterns appear as building blocks for architecting microservices but they are also used to design with other styles like serverless. This is the case of the structural \textbf{API Gateway} (a variation of the Facade pattern), which is usually implemented along with the \textbf{Authorization} pattern. \textbf{Service Discovery} is important considering microservices are organized in the choreography way, rather than orchestration.  The \textbf{Circuit Breaker} for fault-tolerance, and others like \textbf{Tolerant Reader}, \textbf{Broker}, and \textbf{Event-sourcing}. Balalaie et al. \cite{Balalaie2016-yj} exemplify the use of several of these patterns when describing concrete architectures in the context of Spring Cloud Netflix, which integrates the Spring framework with the Netflix OSS project.

Regarding practices and techniques supporting the act of architecting for CD, we identified more general ones like \textbf{container diagrams} and the agile practice of \textbf{simple design} \cite{Chen2018-ih}. Additionally, several techniques like \textbf{Branch by Abstraction}, \textbf{Dependency Injection}, and \textbf{Deployment Configurations} appear to discuss solutions for migration from monolithic to microservices, general service injection and decoupling service logic from its configurations, respectively. Lehmann and Sandnes define three levels of expressiveness for deployment configurations: \textit{``(1) Highly expressive: no highly error-prone manual steps, little learning required; (2) Somewhat expressive: some manual steps; (3) Manual: largely manual deployment, error-prone''} \cite{Lehmann2017-wf}.

\textbf{Consumer-Driven Contracts} and \textbf{Domain-Driven Design} (DDD), including the \textbf{Bounded Context} strategy, are two other practices supporting the goals of architecting for CD. Balalaie et al. mention the benefit of maintaining contracts when handling changes in \cite{Balalaie2016-yj} and Domain-Driven Design is mentioned in \cite{Shahin2016-pp} to ``identify autonomous business capabilities of software architecture'', \textbf{reducing coupling}.

This core phenomenon is contextualized into \textbf{information systems } \cite{Bellomo2013-ey}, \cite{Shahin2019-bb}, \cite{Chen2015-jz}, \cite{Chen2018-ih}, \textbf{large-scale systems} \cite{Bellomo2013-ey}, \cite{Martensson2017-up}, and \textbf{big data projects} \cite{Chen2016-ky}. Such context also involves the presence of \textbf{experienced software developers}, as in all the other three supporting phenomena. 

\subsection{Supporting Operations}

\noindent\fbox{%
\parbox{\textwidth}{%
\textbf{Supporting operations} becomes a critical success factor for enabling Continuous Delivery. \textit{Experienced software developers} need a holistic view of the development process. \textbf{Increased deployability} and \textbf{monitorability} are the main desired effects. To\textbf{ increase deployability}, the main strategies are to \textbf{design environments} that \textbf{maintain parity} and \textbf{reduce set up times}.
\textbf{Experienced software developers} must \textbf{design} their \textbf{environments}, so that \textbf{set up} (and tear down) \textbf{times} are reduced. Environment parity includes \textit{``architecture infrastructure, test architecture, and automation''}. \textit{Maintaining environmental parity} leads to \textbf{increased deployability}.
Finally, \textbf{maintaining the environmental parity} strategy requires \textbf{close collaboration between development and operations} to assure that needs and operational \textbf{requirements} are considered and incorporated into the architecture. Close collaboration can be achieved by giving developers more responsibilities at operations. \textit{Logs} are a key interface for both sides, and the \textbf{collaboration between development and operation} should ensure that \textbf{readable logs} are \textbf{designed} and maintained for improved \textbf{monitorability}, with the consequence of\textbf{ improved issue detection}.
    }%
}

\textbf{Supporting operations} represent the phenomenon that leads organizations to deliver sustainable continuous deployment processes. By applying the strategies associated with this phenomenon, organizations are able to sustain their investments in continuous deployment. As presented in the memo, two main strategies emerged within this phenomenon: \textbf{designing environments} and\textbf{ maintaining environmental parity}.

Environments should be designed to \textbf{reduce setup times}. These reduced setup and teardown times lead directly to \textbf{increased deployability} \cite{Chen2015-jz}. Furthermore, \textbf{maintaining environmental parity} is a strategy which aims at maintaining a reasonable equivalence between the different environments involved in the deployment. We have found evidence that Maintaining environmental parity has the effect of \textbf{increased deployability} \cite{Shahin2019-bb}.

\textbf{Maintaining environmental parity} can be achieved with \textbf{close collaboration between developers} and operations.  These collaborations should enable \textbf{operational requirements} to be \textbf{incorporated into the architecture}, and by giving \textbf{developers more responsibility into operations}  \cite{Shahin2016-pp}.

Finally, system logs are the key interface to achieve the previous strategy. Teams should \textbf{design logs} so that they are \textbf{readable by all stakeholders}. \textbf{Creating readable logs} has the consequence of improving \textbf{monitorability}, which has the immediate positive effect of \textbf{improving issue detection} \cite{Shahin2016-pp}.

\subsection{Continuous Evolution}

\noindent\fbox{%
    \parbox{\textwidth}{%
As organizations seek to increase \textbf{fast delivery capability} and decrease \textbf{cost overrun}, \textbf{continuous evolution} should be a fundamental concern to meet these goals. When implementing continuous evolution, several practices support it, such as \textbf{technical debt management}, \textbf{design memos}, and \textbf{architectural change plan}. Otherwise, the \textbf{continuous evolution} might lead to an excessive \textbf{focus on speed} (ignoring quality attributes), which in turn leads to rapid architectural decay (i.e., \textbf{architectural debt invisible to business}). Another common technique used in this context is \textbf{feature toggle}, even though its usage should be carefully planned since it may increase \textbf{structural complexity}. The formation of the teams is also an important concern for \textbf{continuous evolution} as it requires \textbf{experienced software developers}, which should ideally work in an independent way that, in terms of the \textbf{Conways' Law}. The software architecture can support the teams' autonomy with \textbf{increased} instead of \textbf{decreased deployability}.
    }%
}

\textbf{Continuous evolution} represents the phenomenon that depicts how organizations work with their software systems when adopting CD. We found that in some organizations where the studies were conducted, the CD adoption was motivated by the possibility of reducing \textbf{cost overrun} \cite{Bachmann2012-ik} and improving \textbf{delivery capability} \cite{Bellomo2013-ey}.  

However, these two aspects can translate into an excessive \textbf{focus on speed}, which can bring undesirable consequences, mainly concerned with \textbf{architectural debt} \cite{Bellomo2013-ey}. 
Still, regarding the studies’ context, \textbf{experienced software developers} are important for implementing \textbf{continuous evolution} as they represented the majority of the participants in some studies \cite{Shahin2019-bb}\cite{Martensson2017-up}.

The primary studies report several practices to support the \textbf{continuous evolution} seeking to improve \textbf{delivery capability}. These practices are primarily concerned with internal quality aspects of the system under development. 
In \cite{Bellomo2013-ey}, the authors found that organizations are adopting technical debt management practices, like keeping record of design decisions leading to architectural debt.
These measures are important to keep the system in a maintainable state and support the developers in assessing the impact of system changes over time. 
Also, the authors mention the \textbf{Design Memos}, which represents \textit{``a minimal design document containing architectural design information''}.

Regarding the functional aspects of the system, some practices can be applied as well. In \cite{Shahin2016-pp}, interviewees mentioned and described the use of \textbf{feature toggles} and the \textbf{branch by abstraction pattern} to support incremental changes. Another interesting report related to this issue can be found in \cite{Schermann2018-og}, in which an investigated organization appreciates \textit{``that properly managed and synchronized (e.g., using tools such as ZooKeeper) feature toggles give them more control over their application ecosystem''}. However, as alerted in the previous excerpt, when \textbf{Feature Toggles} are not properly managed and synchronized, technical debt and the additional level of \textbf{structural complexity} emerge \cite{Schermann2018-og}.

Apart from being affected by technical aspects, \textbf{continuous evolution} is also influenced by the team organization. It is noticed in \cite{Shahin2019-bb} that the Conway’s law influenced cross-team dependencies, causing frictions in the CD pipeline. It shows how important the teams’ responsibilities are and how they should be distributed appropriately among the modules of the system under development since. Hence, the team organization and the software architecture match or mismatch can lead to \textbf{increased or decreased deployability}, respectively. 

\subsection{Improving Deployability}

\noindent\fbox{%
    \parbox{\textwidth}{%
\textbf{Experienced software developers} looking at \textbf{improving deployability} should observe how their decisions affect \textbf{availability}, \textbf{deployability}, and \textbf{testability}. To do so, \textbf{experienced software developers} must follow the principles of \textbf{deployability architectural concerns}, \textbf{design for testabilit}y and minimizing \textbf{system deployment dowtimes}. \textbf{Design for testability} has a direct effect on \textbf{testability}. Likewise, minimizing \textbf{system deployment downtime}s has a direct effect on the \textbf{availability} of the software system.
At their disposal lies a toolset of actions that can assist them in \textbf{improving deployability} — a smart and efficient\textbf{ test strategy}, as a direct influence on the capacity to maintain \textbf{efficient deployment pipelines}. Furthermore, \textbf{experienced software developers} can implement several \textbf{Deployability Tactics}, including \textbf{automated deployment},\textbf{ use of cloud resources}, \textbf{gradual deployment strategies}, and \textbf{common platforms}.
    }%
}

\textbf{Improving deployability} represents the phenomenon that is observable when teams of \textbf{experienced software developers} that are \textbf{architecting for CD} start to look at improving the capacity to deploy software. When doing so, they should take into consideration three main quality attributes: \textbf{availability}, \textbf{deployability} and \textbf{testability}.

There are three principles driving to \textbf{improve deployability}. First, the principle of \textbf{deployability architectural concerns} conveys the idea that experienced software developers must take the \textbf{\textit{deployability}} quality attribute into account when designing their arquitectures \cite{Shahin2019-bb}. The second principle of concern is \textbf{designing for testability}. Improving the capacity to deploy a software system imposes challenges to the capacity of the test suite to identify defects. Therefore, the software system must be architected to enable a \textbf{Test Suite} that leverages the tradeoff between thoroughness and performance \cite{Lehmann2017-wf}. Finally, the \textbf{minimize deployment downtime} principle encompasses the drive towards the ideal situation where new deployment should be accomplished with zero downtime. For instance, when mentioning this principle, Chen \cite{Chen2018-ih} goes to the length as affirming that zero downtime is a requirement for Continuous Delivery. This principle has a direct positive effect in \textbf{Availability}.

As a result of these principles, a coherent \textbf{test strategy} is key to \textbf{improve deployability}. It means realising the \textbf{test strategy} is a key factor in the capacity to deploy frequently and continuously. As mentioned, the \textbf{test strategy} must leverage its execution time with its coverages \cite{Shahin2016-pp}. Also, there is a positive relationship that a coherent \textbf{test strategy} has on obtaining \textbf{efficient deployment pipelines} \cite{Chen2018-ih}.

Finally, a set of \textbf{deployability tactics} are at the disposal of \textit{experienced software developers} looking to improve deployability. \textbf{Automated deployments} are a requirement for a sound and consistent capacity to continuously deploy software \cite{Chen2018-ih}. \textbf{Use of cloud resources} is another strategy involving the use of cloud computing infrastructure as the deployment platform \cite{Villamizar2015-hs}. \textbf{Gradual deployment} strategy encompasses the practices that enable experienced software developers to select the components and functionalities to deploy, for instance, the use blue/green deployments, canary release, and gradual rollouts \cite{Schermann2018-og}. Finally, the \textbf{common platform} is the last named strategy in our theory concerning \textbf{improving deployability}. It refers to potential benefits of introducing a common technology stack to benefit from the knowledge of the experienced developers of the software system \cite{Bachmann2012-ik}.

\section{Discussion}

In this section we discuss the results by answering the research questions using the proposed theoretical structure.

\textit{Which architectural characteristics contribute to the delivery capability?}

\textit{RQ1: What are the variables concerning the concrete architecture (factors, mediators, and moderators) influencing the delivery capability?}

The notion of delivery capability can be generally understood as the readiness and competence that teams or organizations have to deliver software, which is usually associated with the desire of delivery frequently. Most of the codes defined as \textit{effects} in the paradigm described in the analysis are related to delivery capability. Nevertheless, among these codes, deployability is a prominent aspect of the study. Two out of four \textit{phenomena} in our analysis have deployability as part of their name (\textbf{architecting for continuous deployment} and \textbf{improving deployability}), including the \textit{core phenomenon}, which indicates the relevance of deployability as an aspect associated with delivery capability.

In short, when looking at the grounded theory (Figure \ref{fig:BigPicture}), the variables influencing aspects related to delivery capability are represented in the theory. Even though we can not claim that our theory is complete, it does convey the diversity of elements concerning the concrete architecture of a system that supports continuous delivery. Based on the developed paradigm, the factors influencing \textit{effects} (\textit{interpreting delivery capability as an effect in the theory}) can be either \textit{context} or \textit{strategies}. For instance, the identified design patterns, architectural styles, deployability tactics, as well as the supporting practices and techniques contribute positively for a concrete architecture with improved deployability and, consequently, an increased delivery capability.

\textit{RQ2 How does the quality of the concrete architecture impact the delivery capability?}

We answer this question focusing on \textbf{deployability} since, as stated in the previous question, it is the most relevant aspect of delivery capability. To answer this question, we interpret the term quality as related to the \textit{Principles} and \textit{Quality Attributes} in the paradigm. In the paradigm (Figure \ref{fig:paradigm}), both constructs provide the motivation (\textbf{why}) for attaining the \textit{Effects}. But these \textit{Effects} can only be attained through the implementation of the \textit{Actions} and \textit{Interactions}. We, therefore, interpret that a concrete architecture of good quality must abide by the principles and conform to the quality attributes.

In the grounded theory (Figure \ref{fig:BigPicture}), this question can be answered by navigating the positive (green connectors) and negative effects (red connectors). Thus, the phenomena that most contribute to this question are \textit{Architecting for Continuous Deployment} and \textit{Continuous Evolution}. \textbf{Increased deployability} is directly related to the \textit{Principles}: \textbf{single responsibility principle} and \textbf{Conway’s law}. \textbf{Single responsibility principle} is crucial for \textbf{testability}, which enables unit testing and other kinds of automated tests that become a requirement when the goal is to deliver continually. Apart from these product concerns, human aspects also impact delivery capability. This is related to \textbf{Conway’s law}, which states that the team's communication structure should mirror the system’s architecture.

Regarding the factors associated with the \textbf{decreased deployability}, one is precisely the result of\textbf{ Conway’s law} potentially negative side. It occurs when the team's communication structure is incompatible with the system’s architecture, creating obstacles to effectively accomplishing the activities planned in the process. Another direct negative influence on \textbf{deployability} is the presence of \textbf{shared databases}. This design is a type of \textbf{tightly coupled architecture} that can impact, for instance, independent deployment of software modules or testing them separately. 

\textit{RQ3: How do these variables relate to each other to contribute to the delivery capability?}

We separate the answer to this question in two levels of abstraction related to the paradigm and the grounded theory that resulted from our analysis. The paradigm described in Figure \ref{fig:paradigm} represents a “high-level” answer to this question. It shows a schema describing the possible ways the codes extracted from the primary studies can relate depending on their types. Although it is not a causal model, it essentially depicts how \textit{context} and \textit{quality attributes} represent motivators or obstacles to achieve the desired \textit{effects} through \textit{strategies} (e.g., \textit{techniques}, \textit{practices}, and \textit{design patterns}) guided by \textit{principles}. Thus, this high-level structure of the paradigm captures the essential elements involved in the connections among the variables and their contribution to the delivery capability. 

Furthermore, the grounded theory in Figure \ref{fig:BigPicture} represents the “low-level” answer as it contains all codes defined during the analysis and their relations following the paradigm. Also, it contains the core phenomenon, which describes how software systems must be architected for CD. As an example, we can identify the mechanism (path from context, quality attributes and principles to effects) in the \textbf{Architecting for Continuous Deployment} phenomenon that can be read as ``driven by \textbf{modifiability} goals, explicitly described in \textbf{design principles} such as the \textbf{single responsibility} as well as the \textbf{small and deployable units}, \textbf{experienced software developers} adopt the \textbf{Microservices} architectural style to facilitate incremental changes.''

\textit{RQ4: How does the effect of the identified variables evolve over time?}

The primary studies included in this investigation do not contain enough elements to provide an answer to this question. However, we described the Continuous Evolution phenomenon. Mostly, it explains the continuous evolution of monolithic architectures in a transition stage to a more flexible approach, so we understand the identified strategies in this phenomenon can support the long-term evolution. Furthermore, we understand this question as a relevant topic to be investigated in future works.

\section{Threats to Validity}

We analyzed the threats to validity according to guidelines in \cite{Petersen2015-if}, \cite{kitchenham2015evidence-based}, but considering the aspects using Maxwell's categorization as in \cite{Petersen2013-uh}.

We handled theoretical validity concerned with searching and selection bias by possibly capturing multiple sources from three search engines and establishing the entire protocol, particularly the inclusion/exclusion criteria,\textit{ a priori}. All the previously known material could be included in the search process (100\% coverage). Additionally, we adopted two search methods to reach relevant data sources: search strings and snowballing. All the concepts achieved in the generated theory are grounded on data from empirical studies and rigorously defined using coding procedures and frequent review of the entire research team.

Publication bias is also a concern. We mitigated it by critically analyzing each term from the referred segments and concepts, comparing against all collected information through the constant comparison method and recurring to dictionaries when no conceptual reference could be obtained.
Regarding descriptive validity, we unbiasedly defined the extraction form and process before the execution to answer the RQs strictly. Apart from the coding procedures, at least one researcher reviewed all information. Also, the definition of a paradigm explaining the phenomena in a metalevel supports the consistent description of each phenomenon. 
Although we identified contextual elements in the theory, it is not possible to claim generalizability as the results miss evidence on several other contexts for software development, which could allow the identification or appearance of new codes and categories. Besides, we achieve no saturation for the developed theory, even using a wide search strategy. Likewise, the limits of explanation of the theory are bounded by the limits of the evidence included in the primary sources. Concepts identified and described in the developed theory remain at the level of abstraction provided by the primary studies. For instance, codes like “coherent test strategy”, “experienced software developers”, and “readable logs” may provide no significant addition to the capacity of explanation, but this is the abstraction level presented in the primary studies. Also, this explains the high number of quotations from \cite{Shahin2019-bb}, as it is the one with more depth in details. This way, we understand additional empirical data should be collected regarding software architecture and continuous deployment for the more abstract parts of the synthesis.

Finally, interpretive validity is achieved when the researchers draw conclusions reasonably given the data. A threat in interpreting the data is researcher bias. For the analysis, we adopted rigorous GT procedures executed in parallel by three researchers to avoid bias and solve inconsistencies. Then, the collective codebook evolved iteratively, remaining only consensual information. In addition, we provide complete traceability from the major categories (phenomena) to the raw data segments and the other way round.

\section{Conclusion}
As businesses change and transform their offerings, software organizations seek to adapt to meet user and client expectations. Continuous Software Engineering methods, techniques, and practices are increasingly being employed in this scenario in which technological innovation and contextual uncertainty are relatively common. While CSE represents a vital tool to deal with this scenario, in this paper, we discussed how the inappropriate focus on architectural concerns could affect delivery capability during the software development process.
We found 14 primary studies regarding this matter. Considering it is a specific aspect of a modern software development context, it is possible to say the research community is at least aware of its importance. On the other hand, it is notably the absence of more robust studies in the set of investigations found in this systematic review. Despite this limitation, this investigation shed light on the mechanisms behind software organizations’ delivery capability from the architectural perspective. The following points were salient in our analysis:
\begin{itemize}
\item Software Architecture plays a crucial role in the capacity to deliver software frequently, especially in the CD context;
    \begin{itemize}
        \item The technical literature is abundant in examples of design patterns and principles related to this subject;
        \item Modularization is an essential aspect of software architecture in CD;
    \end{itemize}
\item Supporting operations should be considered from the beginning, since the software architecture definition, to improve the systems’ monitorability and increase deployability. Moreover, organizations should enable developers to accept more responsibilities at operations;
\item The continuous evolution of software systems is an important aspect when organizations are concerned with increasing delivery capability and avoiding cost overrun. However, it requires particular attention to managing technical debt;
\item Improving deployability is, as might be expected, a major concern for CD. It includes strategies related to deployability such as automation and the use of cloud resources. Also, defining test strategies and minimizing deployments downtime support efficient deployment pipelines and systems’ availability, respectively.
\end{itemize}

We propose the grounded theory as an instrument for researchers and practitioners diagnosing problems and identifying improvement opportunities in controlled studies or real-world scenarios. The theory can be used to identify similar contextual factors, quality attributes, strategies, and effects as described in this paper that support decision-making in other situations or, at least, help explain them. Also, the four phenomena (axial coding categories) described in terms of the open codes offer different perspectives to understand these scenarios and act upon them. For instance, the phenomenon \textbf{supporting operations} can support checking whether the operations are promoting the appropriate strategies (e.g., \textbf{designing logs}) to achieve the desired effects (e.g., \textbf{improved monitorability}).

Still, the theory is far from complete, and further developments are expected and welcome. Some of the hypotheses for further research were enumerated in this work. Others are yet to come based on new studies or expanded understanding of the CD practice and how to architect for it.

\begin{acks}
Authors thank the Brazilian National Council for Scientific and Technological Development (CNPq) under project number 407478/2018-3 for funding this research.
\end{acks}

\bibliographystyle{ACM-Reference-Format}
\bibliography{sample-base}


\begin{thebibliography}{56}


\ifx \showCODEN    \undefined \def \showCODEN     #1{\unskip}     \fi
\ifx \showDOI      \undefined \def \showDOI       #1{#1}\fi
\ifx \showISBNx    \undefined \def \showISBNx     #1{\unskip}     \fi
\ifx \showISBNxiii \undefined \def \showISBNxiii  #1{\unskip}     \fi
\ifx \showISSN     \undefined \def \showISSN      #1{\unskip}     \fi
\ifx \showLCCN     \undefined \def \showLCCN      #1{\unskip}     \fi
\ifx \shownote     \undefined \def \shownote      #1{#1}          \fi
\ifx \showarticletitle \undefined \def \showarticletitle #1{#1}   \fi
\ifx \showURL      \undefined \def \showURL       {\relax}        \fi
\providecommand\bibfield[2]{#2}
\providecommand\bibinfo[2]{#2}
\providecommand\natexlab[1]{#1}
\providecommand\showeprint[2][]{arXiv:#2}

\bibitem[\protect\citeauthoryear{Aleti, Buhnova, Grunske, Koziolek, and
  Meedeniya}{Aleti et~al\mbox{.}}{2013}]%
        {Aleti2013-mh}
\bibfield{author}{\bibinfo{person}{Aldeida Aleti}, \bibinfo{person}{Barbora
  Buhnova}, \bibinfo{person}{Lars Grunske}, \bibinfo{person}{Anne Koziolek},
  {and} \bibinfo{person}{Indika Meedeniya}.} \bibinfo{year}{2013}\natexlab{}.
\newblock \bibinfo{title}{Software Architecture Optimization Methods: A
  Systematic Literature Review}.
\newblock , \bibinfo{numpages}{658--683}~pages.
\newblock


\bibitem[\protect\citeauthoryear{Bachmann, Nord, and Ozakaya}{Bachmann
  et~al\mbox{.}}{2012}]%
        {Bachmann2012-ik}
\bibfield{author}{\bibinfo{person}{Felix Bachmann}, \bibinfo{person}{Robert~L
  Nord}, {and} \bibinfo{person}{Ipek Ozakaya}.}
  \bibinfo{year}{2012}\natexlab{}.
\newblock \showarticletitle{Architectural Tactics to support rapid and agile
  stability}.
\newblock \bibinfo{journal}{\emph{CrossTalk The Journal of Defense Sofwtare
  Engineering}} (\bibinfo{year}{2012}), \bibinfo{pages}{20--25}.
\newblock


\bibitem[\protect\citeauthoryear{Balalaie, Heydarnoori, and Jamshidi}{Balalaie
  et~al\mbox{.}}{2016}]%
        {Balalaie2016-yj}
\bibfield{author}{\bibinfo{person}{A Balalaie}, \bibinfo{person}{A
  Heydarnoori}, {and} \bibinfo{person}{P Jamshidi}.}
  \bibinfo{year}{2016}\natexlab{}.
\newblock \showarticletitle{Microservices Architecture Enables {DevOps}:
  Migration to a {Cloud-Native} Architecture}.
\newblock \bibinfo{journal}{\emph{IEEE Softw.}} \bibinfo{volume}{33},
  \bibinfo{number}{3} (\bibinfo{date}{May} \bibinfo{year}{2016}),
  \bibinfo{pages}{42--52}.
\newblock


\bibitem[\protect\citeauthoryear{Beck}{Beck}{2000}]%
        {Beck2000-pb}
\bibfield{author}{\bibinfo{person}{Kent Beck}.}
  \bibinfo{year}{2000}\natexlab{}.
\newblock \bibinfo{booktitle}{\emph{Extreme Programming Explained: Embrace
  Change}}.
\newblock \bibinfo{publisher}{Addison-Wesley Professional}.
\newblock


\bibitem[\protect\citeauthoryear{{Bellomo}, {Ernst}, {Nord}, and
  {Kazman}}{{Bellomo} et~al\mbox{.}}{2014}]%
        {Bellomo2014-ai}
\bibfield{author}{\bibinfo{person}{S. {Bellomo}}, \bibinfo{person}{N. {Ernst}},
  \bibinfo{person}{R. {Nord}}, {and} \bibinfo{person}{R. {Kazman}}.}
  \bibinfo{year}{2014}\natexlab{}.
\newblock \showarticletitle{Toward Design Decisions to Enable Deployability:
  Empirical Study of Three Projects Reaching for the Continuous Delivery Holy
  Grail}. In \bibinfo{booktitle}{\emph{2014 44th Annual IEEE/IFIP International
  Conference on Dependable Systems and Networks}}. \bibinfo{pages}{702--707}.
\newblock
\urldef\tempurl%
\url{https://doi.org/10.1109/DSN.2014.104}
\showDOI{\tempurl}


\bibitem[\protect\citeauthoryear{Bellomo, Nord, and Ozkaya}{Bellomo
  et~al\mbox{.}}{2013}]%
        {Bellomo2013-ey}
\bibfield{author}{\bibinfo{person}{Stephany Bellomo}, \bibinfo{person}{Robert~L
  Nord}, {and} \bibinfo{person}{Ipek Ozkaya}.} \bibinfo{year}{2013}\natexlab{}.
\newblock \showarticletitle{A study of enabling factors for rapid fielding
  combined practices to balance speed and stability}. In
  \bibinfo{booktitle}{\emph{2013 35th International Conference on Software
  Engineering ({ICSE})}} (San Francisco, CA, USA). \bibinfo{publisher}{IEEE},
  \bibinfo{pages}{982--991}.
\newblock


\bibitem[\protect\citeauthoryear{Bosch}{Bosch}{2014}]%
        {Bosch2014-ag}
\bibfield{editor}{\bibinfo{person}{Jan Bosch}} (Ed.).
  \bibinfo{year}{2014}\natexlab{}.
\newblock \bibinfo{booktitle}{\emph{Continuous Software Engineering}}.
\newblock \bibinfo{publisher}{Springer International Publishing},
  \bibinfo{address}{Cham}.
\newblock


\bibitem[\protect\citeauthoryear{Bosch and Eklund}{Bosch and Eklund}{2012}]%
        {Bosch2012-lk}
\bibfield{author}{\bibinfo{person}{Jan Bosch} {and} \bibinfo{person}{Ulrik
  Eklund}.} \bibinfo{year}{2012}\natexlab{}.
\newblock \showarticletitle{Eternal Embedded Software: Towards Innovation
  Experiment Systems}.
\newblock In \bibinfo{booktitle}{\emph{Leveraging Applications of Formal
  Methods, Verification and Validation. Technologies for Mastering Change}},
  \bibfield{editor}{\bibinfo{person}{Tiziana Margaria} {and}
  \bibinfo{person}{Bernhard Steffen}} (Eds.). \bibinfo{series}{Lecture Notes in
  Computer Science}, Vol.~\bibinfo{volume}{7609}. \bibinfo{publisher}{Springer
  Berlin Heidelberg}, \bibinfo{address}{Berlin, Heidelberg},
  \bibinfo{pages}{19--31}.
\newblock


\bibitem[\protect\citeauthoryear{Carroll and Swatman}{Carroll and
  Swatman}{2000}]%
        {Carroll2000-ng}
\bibfield{author}{\bibinfo{person}{J~M Carroll} {and} \bibinfo{person}{P~A
  Swatman}.} \bibinfo{year}{2000}\natexlab{}.
\newblock \bibinfo{title}{Structured-case: a methodological framework for
  building theory in information systems research}.
\newblock , \bibinfo{numpages}{235--242}~pages.
\newblock


\bibitem[\protect\citeauthoryear{Carver}{Carver}{2007}]%
        {Carver2007-ce}
\bibfield{author}{\bibinfo{person}{Jeffrey Carver}.}
  \bibinfo{year}{2007}\natexlab{}.
\newblock \showarticletitle{The Use of Grounded Theory in Empirical Software
  Engineering}.
\newblock In \bibinfo{booktitle}{\emph{Empirical Software Engineering Issues.
  Critical Assessment and Future Directions: International Workshop, Dagstuhl
  Castle, Germany, June 26-30, 2006. Revised Papers}},
  \bibfield{editor}{\bibinfo{person}{Victor~R Basili}, \bibinfo{person}{Dieter
  Rombach}, \bibinfo{person}{Kurt Schneider}, \bibinfo{person}{Barbara
  Kitchenham}, \bibinfo{person}{Dietmar Pfahl}, {and}
  \bibinfo{person}{Richard~W Selby}} (Eds.). \bibinfo{publisher}{Springer
  Berlin Heidelberg}, \bibinfo{address}{Berlin, Heidelberg},
  \bibinfo{pages}{42--42}.
\newblock


\bibitem[\protect\citeauthoryear{{Chen}, {Kazman}, and {Haziyev}}{{Chen}
  et~al\mbox{.}}{2016}]%
        {Chen2016-ky}
\bibfield{author}{\bibinfo{person}{H. {Chen}}, \bibinfo{person}{R. {Kazman}},
  {and} \bibinfo{person}{S. {Haziyev}}.} \bibinfo{year}{2016}\natexlab{}.
\newblock \showarticletitle{Agile Big Data Analytics Development: An
  Architecture-Centric Approach}. In \bibinfo{booktitle}{\emph{2016 49th Hawaii
  International Conference on System Sciences (HICSS)}}.
  \bibinfo{pages}{5378--5387}.
\newblock
\urldef\tempurl%
\url{https://doi.org/10.1109/HICSS.2016.665}
\showDOI{\tempurl}


\bibitem[\protect\citeauthoryear{Chen}{Chen}{2015}]%
        {Chen2015-jz}
\bibfield{author}{\bibinfo{person}{L Chen}.} \bibinfo{year}{2015}\natexlab{}.
\newblock \showarticletitle{Towards Architecting for Continuous Delivery}. In
  \bibinfo{booktitle}{\emph{2015 12th Working {IEEE/IFIP} Conference on
  Software Architecture}}. \bibinfo{pages}{131--134}.
\newblock


\bibitem[\protect\citeauthoryear{Chen}{Chen}{2018}]%
        {Chen2018-ih}
\bibfield{author}{\bibinfo{person}{Lianping Chen}.}
  \bibinfo{year}{2018}\natexlab{}.
\newblock \showarticletitle{Microservices: Architecting for Continuous Delivery
  and {DevOps}}. \bibinfo{publisher}{IEEE}.
\newblock


\bibitem[\protect\citeauthoryear{Claps, Berntsson~Svensson, and Aurum}{Claps
  et~al\mbox{.}}{2015}]%
        {Claps2015-lj}
\bibfield{author}{\bibinfo{person}{Gerry~Gerard Claps},
  \bibinfo{person}{Richard Berntsson~Svensson}, {and}
  \bibinfo{person}{Ayb{\"u}ke Aurum}.} \bibinfo{year}{2015}\natexlab{}.
\newblock \showarticletitle{On the journey to continuous deployment: Technical
  and social challenges along the way}.
\newblock \bibinfo{journal}{\emph{Information and Software Technology}}
  \bibinfo{volume}{57} (\bibinfo{date}{Jan.} \bibinfo{year}{2015}),
  \bibinfo{pages}{21--31}.
\newblock


\bibitem[\protect\citeauthoryear{Cruzes and Dyb{\aa}}{Cruzes and
  Dyb{\aa}}{2011}]%
        {Cruzes2011-zn}
\bibfield{author}{\bibinfo{person}{Daniela~S Cruzes} {and}
  \bibinfo{person}{Tore Dyb{\aa}}.} \bibinfo{year}{2011}\natexlab{}.
\newblock \bibinfo{title}{Research synthesis in software engineering: A
  tertiary study}.
\newblock , \bibinfo{numpages}{440--455}~pages.
\newblock


\bibitem[\protect\citeauthoryear{de~Fran{\c c}a, Santos, and
  Matalonga}{de~Fran{\c c}a et~al\mbox{.}}{2020}]%
        {De_Franca2020-uy}
\bibfield{author}{\bibinfo{person}{Breno Bernard~Nicolau de Fran{\c c}a},
  \bibinfo{person}{Paulo S{\'e}rgio Medeiros~dos Santos}, {and}
  \bibinfo{person}{Santiago Matalonga}.} \bibinfo{year}{2020}\natexlab{}.
\newblock \bibinfo{booktitle}{\emph{On the relationship between software
  architecture and delivery capability}}.
\newblock \bibinfo{type}{{T}echnical {R}eport} IC-20-07.
  \bibinfo{institution}{Universidade Estadual de Campinas}.
\newblock


\bibitem[\protect\citeauthoryear{Dixon-Woods, Agarwal, Jones, Young, and
  Sutton}{Dixon-Woods et~al\mbox{.}}{2005}]%
        {Dixon-Woods2005-ik}
\bibfield{author}{\bibinfo{person}{Mary Dixon-Woods}, \bibinfo{person}{Shona
  Agarwal}, \bibinfo{person}{David Jones}, \bibinfo{person}{Bridget Young},
  {and} \bibinfo{person}{Alex Sutton}.} \bibinfo{year}{2005}\natexlab{}.
\newblock \showarticletitle{Synthesising qualitative and quantitative evidence:
  a review of possible methods}.
\newblock \bibinfo{journal}{\emph{J. Health Serv. Res. Policy}}
  \bibinfo{volume}{10}, \bibinfo{number}{1} (\bibinfo{date}{Jan.}
  \bibinfo{year}{2005}), \bibinfo{pages}{45--53}.
\newblock


\bibitem[\protect\citeauthoryear{Dyba, Dingsoyr, and Hanssen}{Dyba
  et~al\mbox{.}}{2007}]%
        {Dyba2007-pv}
\bibfield{author}{\bibinfo{person}{Tore Dyba}, \bibinfo{person}{Torgeir
  Dingsoyr}, {and} \bibinfo{person}{Geir~K Hanssen}.}
  \bibinfo{year}{2007}\natexlab{}.
\newblock \bibinfo{title}{Applying Systematic Reviews to Diverse Study Types:
  An Experience Report}.
\newblock
\newblock


\bibitem[\protect\citeauthoryear{Finfgeld-Connett}{Finfgeld-Connett}{2014}]%
        {Finfgeld-Connett2014-fd}
\bibfield{author}{\bibinfo{person}{Deborah Finfgeld-Connett}.}
  \bibinfo{year}{2014}\natexlab{}.
\newblock \bibinfo{title}{Use of content analysis to conduct knowledge-building
  and theory-generating qualitative systematic reviews}.
\newblock , \bibinfo{numpages}{341--352}~pages.
\newblock


\bibitem[\protect\citeauthoryear{Fitzgerald and Stol}{Fitzgerald and
  Stol}{2017}]%
        {Fitzgerald2017-rg}
\bibfield{author}{\bibinfo{person}{Brian Fitzgerald} {and}
  \bibinfo{person}{Klaas-Jan Stol}.} \bibinfo{year}{2017}\natexlab{}.
\newblock \showarticletitle{Continuous software engineering: A roadmap and
  agenda}.
\newblock \bibinfo{journal}{\emph{J. Syst. Softw.}}  \bibinfo{volume}{123}
  (\bibinfo{date}{Jan.} \bibinfo{year}{2017}), \bibinfo{pages}{176--189}.
\newblock


\bibitem[\protect\citeauthoryear{Garousi, Petersen, and Ozkan}{Garousi
  et~al\mbox{.}}{2016}]%
        {Garousi2016-nj}
\bibfield{author}{\bibinfo{person}{Vahid Garousi}, \bibinfo{person}{Kai
  Petersen}, {and} \bibinfo{person}{Baris Ozkan}.}
  \bibinfo{year}{2016}\natexlab{}.
\newblock \bibinfo{title}{Challenges and best practices in industry-academia
  collaborations in software engineering: A systematic literature review}.
\newblock , \bibinfo{numpages}{106--127}~pages.
\newblock


\bibitem[\protect\citeauthoryear{Gay}{Gay}{2011}]%
        {Gay2011-kj}
\bibfield{author}{\bibinfo{person}{Bruce Gay}.}
  \bibinfo{year}{2011}\natexlab{}.
\newblock \bibinfo{title}{Theory building and paradigms: A primer on the
  nuances of theory construction}.
\newblock
  \bibinfo{howpublished}{\url{http://www.aijcrnet.com/journals/Vol_1_No_2_September_2011/4.pdf}}.
\newblock
\newblock
\shownote{Accessed: 2021-1-7.}


\bibitem[\protect\citeauthoryear{Gioia and Pitre}{Gioia and Pitre}{1990}]%
        {Gioia1990-en}
\bibfield{author}{\bibinfo{person}{Dennis~A Gioia} {and}
  \bibinfo{person}{Evelyn Pitre}.} \bibinfo{year}{1990}\natexlab{}.
\newblock \showarticletitle{Multiparadigm Perspectives on Theory Building}.
\newblock \bibinfo{journal}{\emph{The Academy of Management Review}}
  \bibinfo{volume}{15}, \bibinfo{number}{4} (\bibinfo{year}{1990}),
  \bibinfo{pages}{584}.
\newblock


\bibitem[\protect\citeauthoryear{Giustini and Kamel~Boulos}{Giustini and
  Kamel~Boulos}{2013}]%
        {Giustini2013-rw}
\bibfield{author}{\bibinfo{person}{Dean Giustini} {and}
  \bibinfo{person}{Maged~N Kamel~Boulos}.} \bibinfo{year}{2013}\natexlab{}.
\newblock \bibinfo{title}{Google Scholar is not enough to be used alone for
  systematic reviews}.
\newblock
\newblock


\bibitem[\protect\citeauthoryear{Glaser and Strauss}{Glaser and
  Strauss}{2017}]%
        {Glaser2017-bg}
\bibfield{author}{\bibinfo{person}{Barney~G Glaser} {and}
  \bibinfo{person}{Anselm~L Strauss}.} \bibinfo{year}{2017}\natexlab{}.
\newblock \bibinfo{title}{The Discovery of Grounded Theory}.
\newblock , \bibinfo{numpages}{18}~pages.
\newblock


\bibitem[\protect\citeauthoryear{Humble and Farley}{Humble and Farley}{2010}]%
        {Humble2010-rc}
\bibfield{author}{\bibinfo{person}{Jez Humble} {and} \bibinfo{person}{David
  Farley}.} \bibinfo{year}{2010}\natexlab{}.
\newblock \bibinfo{booktitle}{\emph{Continuous Delivery: Reliable Software
  Releases through Build, Test, and Deployment Automation (Adobe Reader)}}.
\newblock \bibinfo{publisher}{Pearson Education}.
\newblock


\bibitem[\protect\citeauthoryear{{ISO}}{{ISO}}{2011}]%
        {ISO_Standard25010}
\bibfield{author}{\bibinfo{person}{{ISO}}.} \bibinfo{year}{2011}\natexlab{}.
\newblock \bibinfo{title}{{ISO/IEC 25010:2011 Systems and software engineering
  -- Systems and software Quality Requirements and Evaluation (SQuaRE) --
  System and software quality models}}.
\newblock , \bibinfo{numpages}{34}~pages.
\newblock


\bibitem[\protect\citeauthoryear{Ivanov and Smolander}{Ivanov and
  Smolander}{2018}]%
        {Ivanov2018-lq}
\bibfield{author}{\bibinfo{person}{Vitalii Ivanov} {and} \bibinfo{person}{Kari
  Smolander}.} \bibinfo{year}{2018}\natexlab{}.
\newblock \showarticletitle{Implementation of a DevOps Pipeline for Serverless
  Applications}. In \bibinfo{booktitle}{\emph{Product-Focused Software Process
  Improvement}}, \bibfield{editor}{\bibinfo{person}{Marco Kuhrmann},
  \bibinfo{person}{Kurt Schneider}, \bibinfo{person}{Dietmar Pfahl},
  \bibinfo{person}{Sousuke Amasaki}, \bibinfo{person}{Marcus Ciolkowski},
  \bibinfo{person}{Regina Hebig}, \bibinfo{person}{Paolo Tell},
  \bibinfo{person}{Jil Kl{\"u}nder}, {and} \bibinfo{person}{Steffen
  K{\"u}pper}} (Eds.). \bibinfo{publisher}{Springer International Publishing},
  \bibinfo{pages}{48--64}.
\newblock
\showISBNx{978-3-030-03673-7}


\bibitem[\protect\citeauthoryear{Kitchenham}{Kitchenham}{2015}]%
        {kitchenham2015evidence-based}
\bibfield{author}{\bibinfo{person}{Barbara Kitchenham}.}
  \bibinfo{year}{2015}\natexlab{}.
\newblock \bibinfo{booktitle}{\emph{Evidence-based Software engineering and
  systematic reviews}}.
\newblock \bibinfo{publisher}{Chapman and Hall/CRC, an imprint of Taylor and
  Francis}, \bibinfo{address}{Boca Raton, FL}.
\newblock
\showISBNx{9780429157653}


\bibitem[\protect\citeauthoryear{Kitchenham and Charters}{Kitchenham and
  Charters}{2007}]%
        {Kitchenham2007-dg}
\bibfield{author}{\bibinfo{person}{Barbara Kitchenham} {and} \bibinfo{person}{B
  Charters}.} \bibinfo{year}{2007}\natexlab{}.
\newblock \bibinfo{booktitle}{\emph{Guidelines for performing Systematic
  Literature Reviews in Software Engineering}}.
\newblock \bibinfo{type}{{T}echnical {R}eport}. \bibinfo{institution}{Keele
  University and Durham University Joint Report}.
\newblock


\bibitem[\protect\citeauthoryear{Kitchenham, Sj{\o}berg, Dyb{\aa}, Pfahl,
  Brereton, Budgen, H{\"o}st, and Runeson}{Kitchenham et~al\mbox{.}}{2012}]%
        {Kitchenham2012-ak}
\bibfield{author}{\bibinfo{person}{Barbara~Ann Kitchenham},
  \bibinfo{person}{Dag I~K Sj{\o}berg}, \bibinfo{person}{Tore Dyb{\aa}},
  \bibinfo{person}{Dietmar Pfahl}, \bibinfo{person}{Pearl Brereton},
  \bibinfo{person}{David Budgen}, \bibinfo{person}{Martin H{\"o}st}, {and}
  \bibinfo{person}{Per Runeson}.} \bibinfo{year}{2012}\natexlab{}.
\newblock \bibinfo{title}{Three empirical studies on the agreement of reviewers
  about the quality of software engineering experiments}.
\newblock , \bibinfo{numpages}{804--819}~pages.
\newblock


\bibitem[\protect\citeauthoryear{Laukkanen, Itkonen, and Lassenius}{Laukkanen
  et~al\mbox{.}}{2017}]%
        {Laukkanen2017-li}
\bibfield{author}{\bibinfo{person}{Eero Laukkanen}, \bibinfo{person}{Juha
  Itkonen}, {and} \bibinfo{person}{Casper Lassenius}.}
  \bibinfo{year}{2017}\natexlab{}.
\newblock \showarticletitle{Problems, causes and solutions when adopting
  continuous delivery---A systematic literature review}.
\newblock \bibinfo{journal}{\emph{Information and Software Technology}}
  \bibinfo{volume}{82} (\bibinfo{date}{Feb.} \bibinfo{year}{2017}),
  \bibinfo{pages}{55--79}.
\newblock


\bibitem[\protect\citeauthoryear{Lehmann and Sandnes}{Lehmann and
  Sandnes}{2017}]%
        {Lehmann2017-wf}
\bibfield{author}{\bibinfo{person}{Martin Lehmann} {and}
  \bibinfo{person}{Frode~Eika Sandnes}.} \bibinfo{year}{2017}\natexlab{}.
\newblock \showarticletitle{A framework for evaluating continuous microservice
  delivery strategies}. In \bibinfo{booktitle}{\emph{Proceedings of the Second
  International Conference on Internet of things, Data and Cloud Computing}}
  (Cambridge, United Kingdom) \emph{(\bibinfo{series}{ICC '17},
  \bibinfo{number}{Article 64})}. \bibinfo{publisher}{Association for Computing
  Machinery}, \bibinfo{address}{New York, NY, USA}, \bibinfo{pages}{1--9}.
\newblock


\bibitem[\protect\citeauthoryear{Leppanen, Makinen, Pagels, Eloranta, Itkonen,
  Mantyla, and Mannisto}{Leppanen et~al\mbox{.}}{2015}]%
        {Leppanen2015-qc}
\bibfield{author}{\bibinfo{person}{Marko Leppanen}, \bibinfo{person}{Simo
  Makinen}, \bibinfo{person}{Max Pagels}, \bibinfo{person}{Veli-Pekka
  Eloranta}, \bibinfo{person}{Juha Itkonen}, \bibinfo{person}{Mika~V Mantyla},
  {and} \bibinfo{person}{Tomi Mannisto}.} \bibinfo{year}{2015}\natexlab{}.
\newblock \showarticletitle{The highways and country roads to continuous
  deployment}.
\newblock \bibinfo{journal}{\emph{IEEE Softw.}} \bibinfo{volume}{32},
  \bibinfo{number}{2} (\bibinfo{date}{March} \bibinfo{year}{2015}),
  \bibinfo{pages}{64--72}.
\newblock


\bibitem[\protect\citeauthoryear{Lynham}{Lynham}{2002}]%
        {Lynham2002-kr}
\bibfield{author}{\bibinfo{person}{Susan~A Lynham}.}
  \bibinfo{year}{2002}\natexlab{}.
\newblock \bibinfo{title}{The General Method of {Theory-Building} Research in
  Applied Disciplines}.
\newblock , \bibinfo{numpages}{221--241}~pages.
\newblock


\bibitem[\protect\citeauthoryear{M{\"a}kinen, Lepp{\"a}nen, Kilamo, Mattila,
  Laukkanen, Pagels, and M{\"a}nnist{\"o}}{M{\"a}kinen et~al\mbox{.}}{2016}]%
        {Makinen2016-vq}
\bibfield{author}{\bibinfo{person}{Simo M{\"a}kinen}, \bibinfo{person}{Marko
  Lepp{\"a}nen}, \bibinfo{person}{Terhi Kilamo}, \bibinfo{person}{Anna-Liisa
  Mattila}, \bibinfo{person}{Eero Laukkanen}, \bibinfo{person}{Max Pagels},
  {and} \bibinfo{person}{Tomi M{\"a}nnist{\"o}}.}
  \bibinfo{year}{2016}\natexlab{}.
\newblock \showarticletitle{Improving the delivery cycle: A multiple-case study
  of the toolchains in Finnish software intensive enterprises}.
\newblock \bibinfo{journal}{\emph{Information and Software Technology}}
  \bibinfo{volume}{80} (\bibinfo{date}{Dec.} \bibinfo{year}{2016}),
  \bibinfo{pages}{175--194}.
\newblock


\bibitem[\protect\citeauthoryear{M{\"a}ntyl{\"a}, Adams, Khomh, Engstr{\"o}m,
  and Petersen}{M{\"a}ntyl{\"a} et~al\mbox{.}}{2015}]%
        {Mantyla2015-fa}
\bibfield{author}{\bibinfo{person}{Mika~V M{\"a}ntyl{\"a}},
  \bibinfo{person}{Bram Adams}, \bibinfo{person}{Foutse Khomh},
  \bibinfo{person}{Emelie Engstr{\"o}m}, {and} \bibinfo{person}{Kai Petersen}.}
  \bibinfo{year}{2015}\natexlab{}.
\newblock \showarticletitle{On rapid releases and software testing: a case
  study and a semi-systematic literature review}.
\newblock \bibinfo{journal}{\emph{Empir. Softw. Eng.}} \bibinfo{volume}{20},
  \bibinfo{number}{5} (\bibinfo{date}{Oct.} \bibinfo{year}{2015}),
  \bibinfo{pages}{1384--1425}.
\newblock


\bibitem[\protect\citeauthoryear{Marschall}{Marschall}{2007}]%
        {Marschall2007-yx}
\bibfield{author}{\bibinfo{person}{Matthias Marschall}.}
  \bibinfo{year}{2007}\natexlab{}.
\newblock \showarticletitle{Transforming a Six Month Release Cycle to
  Continuous Flow}. In \bibinfo{booktitle}{\emph{{AGILE} 2007 ({AGILE} 2007)}}
  (Washington, DC, USA). \bibinfo{publisher}{IEEE}, \bibinfo{pages}{395--400}.
\newblock


\bibitem[\protect\citeauthoryear{{Mårtensson}, {Ståhl}, and
  {Bosch}}{{Mårtensson} et~al\mbox{.}}{2017}]%
        {Martensson2017-up}
\bibfield{author}{\bibinfo{person}{T. {Mårtensson}}, \bibinfo{person}{D.
  {Ståhl}}, {and} \bibinfo{person}{J. {Bosch}}.}
  \bibinfo{year}{2017}\natexlab{}.
\newblock \showarticletitle{Continuous Integration Impediments in Large-Scale
  Industry Projects}. In \bibinfo{booktitle}{\emph{2017 IEEE International
  Conference on Software Architecture (ICSA)}}.
\newblock
\urldef\tempurl%
\url{https://doi.org/10.1109/ICSA.2017.11}
\showDOI{\tempurl}


\bibitem[\protect\citeauthoryear{Olsson, Alahyari, and Bosch}{Olsson
  et~al\mbox{.}}{2012}]%
        {Olsson2012-mx}
\bibfield{author}{\bibinfo{person}{Helena~Holmstrom Olsson},
  \bibinfo{person}{Hiva Alahyari}, {and} \bibinfo{person}{Jan Bosch}.}
  \bibinfo{year}{2012}\natexlab{}.
\newblock \showarticletitle{Climbing the ``Stairway to Heaven'' -- A
  {Mulitiple-Case} Study Exploring Barriers in the Transition from Agile
  Development towards Continuous Deployment of Software}. In
  \bibinfo{booktitle}{\emph{2012 38th Euromicro Conference on Software
  Engineering and Advanced Applications}} (Cesme, Izmir, Turkey).
  \bibinfo{publisher}{IEEE}, \bibinfo{pages}{392--399}.
\newblock


\bibitem[\protect\citeauthoryear{{Petersen} and {Gencel}}{{Petersen} and
  {Gencel}}{2013}]%
        {Petersen2013-uh}
\bibfield{author}{\bibinfo{person}{K. {Petersen}} {and} \bibinfo{person}{C.
  {Gencel}}.} \bibinfo{year}{2013}\natexlab{}.
\newblock \showarticletitle{Worldviews, Research Methods, and their
  Relationship to Validity in Empirical Software Engineering Research}. In
  \bibinfo{booktitle}{\emph{2013 Joint Conference of the 23rd International
  Workshop on Software Measurement and the 8th International Conference on
  Software Process and Product Measurement}}. \bibinfo{pages}{81--89}.
\newblock
\urldef\tempurl%
\url{https://doi.org/10.1109/IWSM-Mensura.2013.22}
\showDOI{\tempurl}


\bibitem[\protect\citeauthoryear{Petersen, Vakkalanka, and Kuzniarz}{Petersen
  et~al\mbox{.}}{2015}]%
        {Petersen2015-if}
\bibfield{author}{\bibinfo{person}{Kai Petersen}, \bibinfo{person}{Sairam
  Vakkalanka}, {and} \bibinfo{person}{Ludwik Kuzniarz}.}
  \bibinfo{year}{2015}\natexlab{}.
\newblock \showarticletitle{Guidelines for conducting systematic mapping
  studies in software engineering: An update}.
\newblock \bibinfo{journal}{\emph{Information and Software Technology}}
  \bibinfo{volume}{64} (\bibinfo{year}{2015}), \bibinfo{pages}{1--18}.
\newblock
\showISSN{0950-5849}
\urldef\tempurl%
\url{https://doi.org/10.1016/j.infsof.2015.03.007}
\showDOI{\tempurl}


\bibitem[\protect\citeauthoryear{Riaz, Sulayman, and Naqvi}{Riaz
  et~al\mbox{.}}{2009}]%
        {Riaz2009-yz}
\bibfield{author}{\bibinfo{person}{Mehwish Riaz}, \bibinfo{person}{Muhammad
  Sulayman}, {and} \bibinfo{person}{Husnain Naqvi}.}
  \bibinfo{year}{2009}\natexlab{}.
\newblock \showarticletitle{Architectural Decay during Continuous Software
  Evolution and Impact of `Design for Change' on Software Architecture}.
\newblock In \bibinfo{booktitle}{\emph{Advances in Software Engineering}},
  \bibfield{editor}{\bibinfo{person}{Dominik {\'S}l{\k e}zak},
  \bibinfo{person}{Tai-Hoon Kim}, \bibinfo{person}{Akingbehin Kiumi},
  \bibinfo{person}{Tao Jiang}, \bibinfo{person}{June Verner}, {and}
  \bibinfo{person}{Silvia Abrah{\~a}o}} (Eds.). \bibinfo{series}{Communications
  in Computer and Information Science}, Vol.~\bibinfo{volume}{59}.
  \bibinfo{publisher}{Springer Berlin Heidelberg}, \bibinfo{address}{Berlin,
  Heidelberg}, \bibinfo{pages}{119--126}.
\newblock


\bibitem[\protect\citeauthoryear{Rodr{\'\i}guez, Haghighatkhah, Lwakatare,
  Teppola, Suomalainen, Eskeli, Karvonen, Kuvaja, Verner, and
  Oivo}{Rodr{\'\i}guez et~al\mbox{.}}{2017}]%
        {Rodriguez2017-wg}
\bibfield{author}{\bibinfo{person}{Pilar Rodr{\'\i}guez},
  \bibinfo{person}{Alireza Haghighatkhah}, \bibinfo{person}{Lucy~Ellen
  Lwakatare}, \bibinfo{person}{Susanna Teppola}, \bibinfo{person}{Tanja
  Suomalainen}, \bibinfo{person}{Juho Eskeli}, \bibinfo{person}{Teemu
  Karvonen}, \bibinfo{person}{Pasi Kuvaja}, \bibinfo{person}{June~M Verner},
  {and} \bibinfo{person}{Markku Oivo}.} \bibinfo{year}{2017}\natexlab{}.
\newblock \showarticletitle{Continuous deployment of software intensive
  products and services: A systematic mapping study}.
\newblock \bibinfo{journal}{\emph{J. Syst. Softw.}}  \bibinfo{volume}{123}
  (\bibinfo{date}{Jan.} \bibinfo{year}{2017}), \bibinfo{pages}{263--291}.
\newblock


\bibitem[\protect\citeauthoryear{Runeson and H{\"o}st}{Runeson and
  H{\"o}st}{2009}]%
        {Runeson2009-sg}
\bibfield{author}{\bibinfo{person}{Per Runeson} {and} \bibinfo{person}{Martin
  H{\"o}st}.} \bibinfo{year}{2009}\natexlab{}.
\newblock \bibinfo{title}{Guidelines for conducting and reporting case study
  research in software engineering}.
\newblock , \bibinfo{numpages}{131--164}~pages.
\newblock


\bibitem[\protect\citeauthoryear{Schermann, Cito, Leitner, Zdun, and
  Gall}{Schermann et~al\mbox{.}}{2018}]%
        {Schermann2018-og}
\bibfield{author}{\bibinfo{person}{Gerald Schermann}, \bibinfo{person}{Jürgen
  Cito}, \bibinfo{person}{Philipp Leitner}, \bibinfo{person}{Uwe Zdun}, {and}
  \bibinfo{person}{Harald~C. Gall}.} \bibinfo{year}{2018}\natexlab{}.
\newblock \showarticletitle{We’re doing it live: A multi-method empirical
  study on continuous experimentation}.
\newblock \bibinfo{journal}{\emph{Information and Software Technology}}
  \bibinfo{volume}{99} (\bibinfo{year}{2018}), \bibinfo{pages}{41--57}.
\newblock
\showISSN{0950-5849}
\urldef\tempurl%
\url{https://doi.org/10.1016/j.infsof.2018.02.010}
\showDOI{\tempurl}


\bibitem[\protect\citeauthoryear{Shahin, Ali~Babar, and Zhu}{Shahin
  et~al\mbox{.}}{2017}]%
        {Shahin2017-yi}
\bibfield{author}{\bibinfo{person}{M Shahin}, \bibinfo{person}{M Ali~Babar},
  {and} \bibinfo{person}{L Zhu}.} \bibinfo{year}{2017}\natexlab{}.
\newblock \showarticletitle{Continuous Integration, Delivery and Deployment: A
  Systematic Review on Approaches, Tools, Challenges and Practices}.
\newblock \bibinfo{journal}{\emph{IEEE Access}}  \bibinfo{volume}{5}
  (\bibinfo{year}{2017}), \bibinfo{pages}{3909--3943}.
\newblock


\bibitem[\protect\citeauthoryear{Shahin, Babar, and Zhu}{Shahin
  et~al\mbox{.}}{2016}]%
        {Shahin2016-pp}
\bibfield{author}{\bibinfo{person}{Mojtaba Shahin},
  \bibinfo{person}{Muhammad~Ali Babar}, {and} \bibinfo{person}{Liming Zhu}.}
  \bibinfo{year}{2016}\natexlab{}.
\newblock \showarticletitle{The Intersection of Continuous Deployment and
  Architecting Process: Practitioners' Perspectives}. In
  \bibinfo{booktitle}{\emph{Proceedings of the 10th {ACM/IEEE} International
  Symposium on Empirical Software Engineering and Measurement}} (Ciudad Real,
  Spain) \emph{(\bibinfo{series}{ESEM '16}, \bibinfo{number}{Article 44})}.
  \bibinfo{publisher}{Association for Computing Machinery},
  \bibinfo{address}{New York, NY, USA}, \bibinfo{pages}{1--10}.
\newblock


\bibitem[\protect\citeauthoryear{Shahin, Zahedi, Babar, and Zhu}{Shahin
  et~al\mbox{.}}{2019}]%
        {Shahin2019-bb}
\bibfield{author}{\bibinfo{person}{Mojtaba Shahin}, \bibinfo{person}{Mansooreh
  Zahedi}, \bibinfo{person}{Muhammad~Ali Babar}, {and} \bibinfo{person}{Liming
  Zhu}.} \bibinfo{year}{2019}\natexlab{}.
\newblock \showarticletitle{An empirical study of architecting for continuous
  delivery and deployment}.
\newblock \bibinfo{journal}{\emph{Empir. Softw. Eng.}} \bibinfo{volume}{24},
  \bibinfo{number}{3} (\bibinfo{date}{June} \bibinfo{year}{2019}),
  \bibinfo{pages}{1061--1108}.
\newblock


\bibitem[\protect\citeauthoryear{Stol, Ralph, and Fitzgerald}{Stol
  et~al\mbox{.}}{2016}]%
        {Stol2016-qn}
\bibfield{author}{\bibinfo{person}{Klaas-Jan Stol}, \bibinfo{person}{Paul
  Ralph}, {and} \bibinfo{person}{Brian Fitzgerald}.}
  \bibinfo{year}{2016}\natexlab{}.
\newblock \showarticletitle{Grounded theory in software engineering research: a
  critical review and guidelines}. In \bibinfo{booktitle}{\emph{Proceedings of
  the 38th International Conference on Software Engineering}} (Austin, Texas)
  \emph{(\bibinfo{series}{ICSE '16})}. \bibinfo{publisher}{Association for
  Computing Machinery}, \bibinfo{address}{New York, NY, USA},
  \bibinfo{pages}{120--131}.
\newblock


\bibitem[\protect\citeauthoryear{Straus and Corbin}{Straus and Corbin}{1990}]%
        {Straus1990-oz}
\bibfield{author}{\bibinfo{person}{Anselm Straus} {and} \bibinfo{person}{Juliet
  Corbin}.} \bibinfo{year}{1990}\natexlab{}.
\newblock \bibinfo{title}{Basics of qualitative research: Grounded theory
  procedures and techniques}.
\newblock
\newblock


\bibitem[\protect\citeauthoryear{Villamizar, Garces, Castro, Verano, Salamanca,
  Casallas, and Gil}{Villamizar et~al\mbox{.}}{2015}]%
        {Villamizar2015-hs}
\bibfield{author}{\bibinfo{person}{Mario Villamizar}, \bibinfo{person}{Oscar
  Garces}, \bibinfo{person}{Harold Castro}, \bibinfo{person}{Mauricio Verano},
  \bibinfo{person}{Lorena Salamanca}, \bibinfo{person}{Rubby Casallas}, {and}
  \bibinfo{person}{Santiago Gil}.} \bibinfo{year}{2015}\natexlab{}.
\newblock \showarticletitle{Evaluating the monolithic and the microservice
  architecture pattern to deploy web applications in the cloud}. In
  \bibinfo{booktitle}{\emph{2015 10th Computing Colombian Conference
  ({10CCC})}} (Bogota, Colombia). \bibinfo{publisher}{IEEE},
  \bibinfo{pages}{583--590}.
\newblock


\bibitem[\protect\citeauthoryear{Virmani}{Virmani}{2015}]%
        {Virmani2015-jn}
\bibfield{author}{\bibinfo{person}{M Virmani}.}
  \bibinfo{year}{2015}\natexlab{}.
\newblock \showarticletitle{Understanding {DevOps} bridging the gap from
  continuous integration to continuous delivery}. In
  \bibinfo{booktitle}{\emph{Fifth International Conference on the Innovative
  Computing Technology ({INTECH} 2015)}}. \bibinfo{pages}{78--82}.
\newblock


\bibitem[\protect\citeauthoryear{Wohlin}{Wohlin}{2014}]%
        {Wohlin2014-bf}
\bibfield{author}{\bibinfo{person}{Claes Wohlin}.}
  \bibinfo{year}{2014}\natexlab{}.
\newblock \showarticletitle{Guidelines for snowballing in systematic literature
  studies and a replication in software engineering}. In
  \bibinfo{booktitle}{\emph{Proceedings of the 18th International Conference on
  Evaluation and Assessment in Software Engineering - {EASE} '14}} (London,
  England, United Kingdom). \bibinfo{publisher}{ACM Press},
  \bibinfo{address}{New York, New York, USA}, \bibinfo{pages}{1--10}.
\newblock


\bibitem[\protect\citeauthoryear{Wolfswinkel, Furtmueller, and
  Wilderom}{Wolfswinkel et~al\mbox{.}}{2013}]%
        {Wolfswinkel2013-fn}
\bibfield{author}{\bibinfo{person}{Joost~F Wolfswinkel}, \bibinfo{person}{Elfi
  Furtmueller}, {and} \bibinfo{person}{Celeste P~M Wilderom}.}
  \bibinfo{year}{2013}\natexlab{}.
\newblock \bibinfo{title}{Using grounded theory as a method for rigorously
  reviewing literature}.
\newblock , \bibinfo{numpages}{45--55}~pages.
\newblock


\bibitem[\protect\citeauthoryear{Zhu, Bass, and Champlin-Scharff}{Zhu
  et~al\mbox{.}}{2016}]%
        {Zhu2016-hu}
\bibfield{author}{\bibinfo{person}{L Zhu}, \bibinfo{person}{L Bass}, {and}
  \bibinfo{person}{G Champlin-Scharff}.} \bibinfo{year}{2016}\natexlab{}.
\newblock \showarticletitle{{DevOps} and Its Practices}.
\newblock \bibinfo{journal}{\emph{IEEE Softw.}} \bibinfo{volume}{33},
  \bibinfo{number}{3} (\bibinfo{date}{May} \bibinfo{year}{2016}),
  \bibinfo{pages}{32--34}.
\newblock


\end{thebibliography}

\clearpage

\end{document}